\documentclass[conference,compsoc]{IEEEtran}

\usepackage{silence}
\usepackage[hyphens]{url}
\usepackage{xurl}
\usepackage{soul}
\usepackage[dvipsnames,table]{xcolor}
\usepackage[breaklinks,colorlinks]{hyperref}
\hypersetup{citecolor=blue,linkcolor=blue}
\usepackage{amsmath,amsopn,amssymb}
\usepackage{subfigure}
\usepackage{endnotes,xspace,graphicx,fancyvrb,multirow}
\usepackage{booktabs}
\usepackage{threeparttable}
\usepackage{array,underscore,relsize}
\usepackage{enumitem}
\usepackage[labelfont=bf,font=small,skip=5pt]{caption}

\usepackage{fp}
\usepackage{siunitx}

\usepackage{minted}

\usepackage[ruled,vlined,linesnumbered]{algorithm2e}

\usepackage{pifont}

\usepackage{balance}

\usepackage{fixltx2e}

\usepackage{verbatim}

\usepackage[most]{tcolorbox}

\newcommand{\sys}{\mbox{\textsc{Gondar}}\xspace}
\newcommand{\sysGPT}{\mbox{\textsc{Gondar}\textsubscript{gpt-5}}\xspace}
\newcommand{\sysGemini}{\mbox{\textsc{Gondar}\textsubscript{gemini-2.5-pro}}\xspace}
\newcommand{\sysSonnet}{\mbox{\textsc{Gondar}\textsubscript{sonnet-4.5}}\xspace}
\newcommand{\sysGLM}{\mbox{\textsc{Gondar}\textsubscript{glm-5}}\xspace}
\newcommand{\sysFlashLite}{\mbox{\textsc{Gondar}\textsubscript{flash-lite}}\xspace}
\newcommand{\sysNano}{\mbox{\textsc{Gondar}\textsubscript{gpt-5-nano}}\xspace}
\newcommand{\sysGPTOSS}{\mbox{\textsc{Gondar}\textsubscript{gpt-oss-120b}}\xspace}

\newcommand{\GsysAF}{\mbox{\textsc{G}-AF}}
\newcommand{\GsysRO}{\mbox{\textsc{G}-RO}}
\newcommand{\GsysXO}{\mbox{\textsc{G}-XO}}
\newcommand{\GsysGPT}{\mbox{\textsc{G}\textsubscript{gpt-5}}}
\newcommand{\GsysGemini}{\mbox{\textsc{G}\textsubscript{gem-2.5-pro}}}
\newcommand{\GsysSonnet}{\mbox{\textsc{G}\textsubscript{sonnet-4.5}}}
\newcommand{\GsysGLM}{\mbox{\textsc{G}\textsubscript{glm-5}}}
\newcommand{\GsysFlashLite}{\mbox{\textsc{G}\textsubscript{flash-lite}}}
\newcommand{\GsysNano}{\mbox{\textsc{G}\textsubscript{gpt-5-nano}}}
\newcommand{\GsysGPTOSS}{\mbox{\textsc{G}\textsubscript{gpt-oss-120b}}}

\newcommand{\baseline}{Jazzer\xspace}
\newcommand{\baselineLS}{Jazzer-LS\xspace}
\newcommand{\directedJazzer}{Directed Jazzer\xspace}
\newcommand{\reactPro}{ReAct\textsubscript{gemini-2.5-pro}\xspace}
\newcommand{\reactFL}{ReAct\textsubscript{flash-lite}\xspace}

\newcommand{\cc}[1]{\mbox{\smaller[0.5]\texttt{#1}}}

\newcommand{\numprojects}{22\xspace}            
\newcommand{\numharnesses}{53\xspace}           
\newcommand{\numcpvs}{54\xspace}                
\newcommand{\numcpvsCVEs}{19\xspace}            
\newcommand{\numcpvsCVEsCWEs}{6\xspace}         
\newcommand{\numcpvsAIxCC}{20\xspace}           
\newcommand{\numcpvsSynthetic}{15\xspace}       

\newcommand{\numcwes}{12\xspace}                

\newcommand{\numloclines}{14k\xspace}                


\fvset{fontsize=\scriptsize,xleftmargin=8pt,numbers=left,numbersep=5pt,breaklines=true,breakanywhere=true}

\definecolor{codehighlightgrey}{gray}{0.9}

\input{code/fmt}

\def\Snospace~{\S{}}

\newcommand{\x}{$\times$\xspace}

\newif\ifdraft\drafttrue
\newif\ifnotes\notestrue
\ifdraft\else\notesfalse\fi

\input{glyphtounicode}
\newcolumntype{R}[1]{>{\raggedleft\let\newline\\\arraybackslash\hspace{0pt}}p{#1}}

\newcommand{\includepdf}[1]{
  \includegraphics[width=\columnwidth]{#1}
}

\newcommand{\squishlist}{
\begin{itemize}[noitemsep,nolistsep]
  \setlength{\itemsep}{-0pt}
}
\newcommand{\squishend}{
  \end{itemize}
}

\usepackage{tikz}
\usetikzlibrary{calc,backgrounds}

\makeatletter
\sethlcolor{codehighlightgrey}

\makeatother

\usepackage{xstring}
\newcommand{\PP}[1]{
\vspace{2px}
\noindent{\bf \IfEndWith{#1}{.}{#1}{#1.}}
}

\newcommand{\PN}[1]{
\vspace{2px}
\noindent{\bf #1}
}

\newcommand{\MB}{\,\text{MB}\xspace}
\newcommand{\GB}{\,\text{GB}\xspace}

\newcommand{\USD}{\,\text{USD}\xspace}

\newcommand{\boxbeg}{%
\vspace{2px}%
\noindent\begin{tabular}{|l|}\hline
\begin{minipage}{\columnwidth}%
\vspace{2px}%
\noindent
}

\newcommand{\boxend}{%
\vspace{2px}%
\end{minipage}\\ \hline
\end{tabular}%
\vspace{-10pt}%
}

\newcounter{findingctr}
\newtcolorbox{findingbox}{%
  enhanced,
  colback=gray!8,
  colframe=gray!60,
  boxrule=0pt,
  leftrule=2.5pt,
  sharp corners,
  left=4pt, right=4pt, top=1pt, bottom=1pt,
  before skip=6pt, after skip=6pt,
  fontupper=\small,
}
\newcommand{\finding}[1]{%
\stepcounter{findingctr}%
\begin{findingbox}%
\textbf{Takeaway~\arabic{findingctr}:} #1%
\end{findingbox}%
}

\gdef\therev{76b1718}
\gdef\thedate{2026-04-11 11:18:07 -0400}

\begin{document}

\title{Contextualizing Sink Knowledge for Java Vulnerability Discovery}

\ifdefined\DRAFT
 \pagestyle{fancyplain}
 \lhead{Rev.~\therev}
 \rhead{\thedate}
 \cfoot{\thepage\ of \pageref{LastPage}}
\fi

\author{
  \IEEEauthorblockN{%
  Fabian Fleischer\IEEEauthorrefmark{2}\IEEEauthorrefmark{1}\quad
  Cen Zhang\IEEEauthorrefmark{2}\IEEEauthorrefmark{1}\quad
  Joonun Jang\IEEEauthorrefmark{3}\quad
  Jeongin Cho\IEEEauthorrefmark{3}\\
  Meng Xu\IEEEauthorrefmark{4}\quad
  Taesoo Kim\IEEEauthorrefmark{2}\IEEEauthorrefmark{3}}
  \IEEEauthorblockA{\IEEEauthorrefmark{2}Georgia Institute of Technology\quad
  \IEEEauthorrefmark{3}Samsung Research\quad
  \IEEEauthorrefmark{4}University of Waterloo}
  \IEEEauthorblockA{
    \texttt{fleischer@gatech.edu},
    \texttt{blbllhy@gmail.com},
    \texttt{joonun.jang@gmail.com},\\
    \texttt{potress7@gmail.com},
    \texttt{meng.xu.cs@uwaterloo.ca},
    \texttt{taesoo@gatech.edu}
  }
}

\date{}
\maketitle
\makeatletter
\let\savemakefntext\@makefntext
\long\def\@makefntext#1{#1}
\footnotetext{\IEEEauthorrefmark{1}\,These authors contributed equally to this work.}
\let\@makefntext\savemakefntext
\makeatother

\sloppy

\begin{abstract}
Java applications are prone to vulnerabilities stemming from the insecure use of security-sensitive APIs, such as file operations enabling path traversal or deserialization routines allowing remote code execution.
These sink APIs encode critical information for vulnerability discovery: the program-specific constraints required to reach them and the exploitation conditions necessary to trigger security flaws.
Despite this, existing fuzzers largely overlook such vulnerability-specific knowledge, limiting their effectiveness.

We present \sys, a sink-centric fuzzing framework that systematically leverages sink API semantics for targeted vulnerability discovery.
\sys first identifies reachable and exploitable sink call sites through CWE-specific scanning combined with LLM-assisted static filtering.
It then deploys two specialized agents that work collaboratively with a coverage-guided fuzzer: an exploration agent generates inputs to reach target call sites by iteratively solving path constraints, while an exploitation agent synthesizes proof-of-concept exploits by reasoning about and satisfying vulnerability-triggering conditions.
The agents and fuzzer continuously exchange seeds and runtime feedback, complementing each other.
We evaluated \sys on real-world Java benchmarks, where it discovers four times more vulnerabilities than Jazzer, the state-of-the-art Java fuzzer.
Notably, an earlier \sys version contributed to Team Atlanta's first-place CRS in the DARPA AI Cyber Challenge, and is integrated into OSS-CRS, a sandbox project in The Linux Foundation's OpenSSF, to analyze open-source Java projects, where it has already uncovered a zero-day vulnerability.
\end{abstract}

\section{Introduction}
\label{s:intro}

Java applications power critical enterprise infrastructure worldwide,
from banking systems and healthcare platforms to e-commerce services and cloud computing.
However,
this widespread adoption makes Java vulnerabilities particularly impactful.
The 2021 Log4Shell vulnerability in Apache Log4j,
for example,
affected hundreds of millions of devices
and was characterized by CISA as one of the most serious security incidents in recent history~\cite{log4j-cisa}.
Many such vulnerabilities manifest
at security-sensitive API call sites, known as sinks,
where untrusted inputs can trigger unsafe operations.
For instance,
a command injection vulnerability occurs when
user-controlled input flows to a system command execution API like \cc{Runtime.exec()}
without proper sanitization,
allowing attackers to execute arbitrary commands on the server.

Effectively discovering these vulnerabilities requires
extracting and utilizing sink-specific knowledge,
the contextual information surrounding each sink
that determines both reachability and exploitability.
This knowledge spans multiple dimensions:
program context captures how execution reaches the sink
through control flow paths, input validation logic, and data dependencies;
API semantics defines what constitutes unsafe usage
in terms of parameter constraints, argument types, and behavioral specifications;
and vulnerability characteristics specify exploitation conditions
including CWE-specific requirements and sanitizer triggers.
Turning this multi-faceted sink knowledge into actionable and automated testing strategies
demands both structural and semantic understanding of program behavior.

Existing dynamic testing approaches fall short
in systematically leveraging sink knowledge.
Coverage-guided fuzzers such as Jazzer~\cite{jazzer} and JQF~\cite{jqf}
inherit their design from C/C++ memory corruption testing,
treating all code paths equally without prioritizing security-sensitive sinks.
CWE-specific approaches address individual vulnerability types
through specialized, target-particular techniques~\cite{fugio,oddfuzz,ssrf-discovery,ssti-rce,gadget-search,jaex},
where cross-CWE generalization is a secondary consideration by design.
While some works have proposed general fuzzing frameworks with sink-awareness,
such as WDFuzz~\cite{wdfuzz}, Witcher~\cite{witcher}, Atropos~\cite{atropos}, and Predator~\cite{predator},
these approaches rely solely on traditional program analysis techniques
like taint tracking and directed scheduling,
which limits the sink knowledge that can be effectively utilized.

In short,
there remains a need for a general and scalable approach
that systematically leverages both structural and semantic sink knowledge
to identify, reach, and exploit security-sensitive sinks across a diverse and growing set of vulnerability types in Java programs.

To address this gap,
we present \sys,
a sink-centric fuzzing framework
that addresses this challenge
through collaborative integration of LLMs' semantic reasoning,
program analysis' structural understanding,
and fuzzing's dynamic exploration.
\sys decomposes sink-based vulnerability discovery
into the two sub-tasks of reachability and exploitability,
and designs agents with task-specific LLM reasoning strategies
for each,
while treating agents and fuzzer
as mutually beneficial collaborators
rather than independent tools.
Given a Java program and a CWE category of interest,
\sys orchestrates three core components.
First, a sink detection component identifies high-potential sinks within the target program
by extracting candidates from CodeQL's CWE-specific sink database~\cite{codeql}
and refining them through multi-dimensional filtering:
validity checks eliminate sinks with constant arguments or in test code,
reachability analysis filters unreachable locations,
and LLM-based exploitability assessment focuses efforts on truly vulnerable targets.
Second, a sink exploration agent generates inputs to reach identified sinks.
The agent operates on the insight that
call paths from program entry to sink naturally encode the fundamental reachability constraints.
By examining code context along these paths,
the agent reasons about input format requirements, branching conditions, and API semantics
to synthesize inputs that progressively satisfy path constraints.
Third, a sink exploitation agent receives sink-reaching inputs from the fuzzer
and develops proof-of-concept exploits.
These inputs provide concrete execution context, such as stack traces, program state, and target sink exploitation specifics, that
ground the agent's reasoning in actual exploit development issues rather than hypothetical scenarios.
The agent iteratively refines exploit attempts
based on execution feedback,
leveraging vulnerability-specific knowledge to satisfy sanitizer conditions.
Throughout the fuzzing campaign,
these agents operate concurrently with the fuzzer,
exchanging information bidirectionally:
agent-generated inputs enrich the fuzzer's corpus for mutation-based exploration,
while fuzzer-discovered sink-reaching inputs provide concrete starting points for exploitation.
Eventually, the fuzzer reports any successful vulnerability discovery as sanitizer violations.
This collaborative pipeline enables \sys
to systematically leverage sink knowledge across diverse vulnerability types.

We evaluate \sys on a new benchmark
comprising \numcpvs vulnerabilities across \numprojects projects
spanning \numcwes CWE types.
\sys significantly outperforms the state-of-the-art Java fuzzer Jazzer,
exploiting 41 vulnerabilities compared to Jazzer's eight
(a 4\x improvement)
and reaching 46 vulnerabilities versus 26
(a 77\% improvement).
Ablation studies confirm
that both sink exploration and exploitation agents are critical:
disabling exploration reduces reached vulnerabilities by 31\%,
while disabling exploitation cuts exploited vulnerabilities by 51\%.
Furthermore,
the synergy between fuzzing and agents
proves essential,
as their collaboration discovers seven additional vulnerabilities
that neither approach can find independently.
Notably,
\sys achieves these results
at lower cost than large-scale fuzzing
(\$3,051 vs. \$3,264)
while being substantially more effective.

Beyond our benchmark,
\sys's techniques have been validated in real-world deployments.
DARPA invested tens of thousands of dollars
in the AI Cyber Challenge (AIxCC),
where an early version of \sys's techniques
contributed to Team Atlanta's first-place CRS
and discovered seven synthetic and three zero-day vulnerabilities in real-world Java projects.
Following these results,
the Open Source Security Foundation (OpenSSF)
reached out for collaboration,
leading to \sys's integration
into OSS-CRS~\cite{oss-crs}, a sandbox project in the OpenSSF,
to continuously protect open-source software.

\PP{Contributions} Our contributions are:
\squishlist
\item A cross-CWE sink-centric fuzzing framework
that deeply integrates LLMs with fuzzing
to systematically leverage sink knowledge.

\item An implementation of \sys supporting \numcwes CWE types
with full OSS-Fuzz compatibility.

\item A benchmark
of \numcpvs vulnerabilities across \numprojects projects
demonstrating significant improvements
in vulnerability discovery ability of \sys.

\item Real-world deployment: an earlier \sys version
contributed the most detected Java vulnerabilities
to Team Atlanta's first-place CRS
in the DARPA AIxCC,
and \sys is now integrated into OSS-CRS~\cite{oss-crs}
in the OpenSSF.
\squishend
We published our prototype, benchmark, and evaluation results
(see \autoref{s:artifact} for details).

\section{Overview}
\label{s:prelim}

\subsection{Background}

\PP{Sinks and Beep Seeds}
A \emph{sink} is a program location
that performs security-sensitive operations,
such as executing system commands,
handling file I/O, or processing network data.
A \emph{sink API} is the specific function or method call at a sink location.
For a given vulnerability type,
the set of sink APIs is extensible:
beyond language-provided primitives like \cc{Runtime.exec()},
third-party libraries that wrap these primitives
can also serve as sink APIs.
A \emph{beep seed} is a user-controllable input
that leads program execution to reach a sink.
While this indicates potential vulnerability exposure,
it does not necessarily trigger an exploit,
as the execution may be benign.

\PP{Sink-Based Java Vulnerabilities}
Many common Java vulnerabilities root in unsafe usage of sink APIs.
These sink-based vulnerabilities occur
when untrusted inputs can exploit the sink usage to trigger unsafe operations.
This paper focuses on prevalent vulnerability types in our dataset,
including command injection,
SQL injection,
path traversal,
XML external entity (XXE) attacks,
deserialization vulnerabilities,
code injection,
XPath injection,
and server-side request forgery (SSRF), etc.
While this coverage is not exhaustive,
our approach generalizes to any sink-based vulnerability
that can be detected through dynamic analysis.

\PP{Jazzer and Its Sanitizers}
Jazzer is a coverage-guided fuzzer for Java built on libFuzzer's foundation.
It inherits libFuzzer's mutation strategies and seed scheduling mechanisms,
while adding a Java-specific layer
that provides bytecode instrumentation for coverage feedback,
custom mutators, and other enhancements.
Beyond detecting vulnerabilities through uncaught exceptions and non-terminations,
Jazzer implements sanitizers for the sink-based vulnerabilities discussed above~\cite{jazzer-sanitizers}.
These sanitizers hook sink APIs in Java's standard library
and provide value profile feedback~\cite{libfuzzer-value-profile}
to guide fuzzing toward exploitation when execution reaches these sinks.

\subsection{Limitations of Coverage-Guided Fuzzing}

To understand the current limitations
of coverage-guided fuzzing on Java vulnerability detection,
we conducted large-scale fuzzing experiments to assess Jazzer's effectiveness
on our Java vulnerability dataset.

\PP{Experiment Setup}
We built a dataset containing \numcpvs vulnerabilities
across \numharnesses fuzzing harnesses from \numprojects projects
(detailed in \autoref{s:eval}).
For each harness,
we deployed 50 fuzzing instances,
each running on one CPU core for 24 hours.
We used Jazzer as the fuzzing engine with default settings.
Input seeds were sourced from OSS-Fuzz~\cite{oss-fuzz}
when harnesses and corpora existed there;
otherwise, we started with empty seeds.
The total computation budget amounted to over \num{7.2} CPU-years (or \num{63600} CPU-hours).

\begin{table}[t]
\centering
\caption{Large-scale fuzzing results
showing vulnerable sinks (one per vulnerability) reached and exploited by Jazzer.
}
\label{tab:large-scale-fuzzing-results}
\scriptsize
\begin{tabular}{lccc}
\toprule
\textbf{Total} & \textbf{Not Reached} & \textbf{Reached Only} & \textbf{Exploited} \\
\midrule
54 & 25 (46.3\%) & 21 (38.9\%) & 8 (14.8\%) \\
\bottomrule
\end{tabular}
\end{table}

\PP{Results and Analysis}
\autoref{tab:large-scale-fuzzing-results} presents the results.
Jazzer successfully exploited only 14.8\% of the existing vulnerabilities,
leaving substantial room for improvement.
The results reveal two distinct failure modes:
46.3\% of vulnerabilities were never reached,
meaning the fuzzer could not generate inputs
to trigger execution of their vulnerable sinks.
More notably,
38.9\% of vulnerabilities had their sinks reached
but remained unexploited, i.e., the
fuzzer failed to craft inputs
that satisfy the sanitizers' exploitation conditions.
We refer to this phenomenon as the ``last mile challenge.''

These results demonstrate that
sink reachability and exploitability
represent fundamentally distinct challenges.
While Jazzer's continuous coverage exploration
enables it to discover many sinks,
its value profile-based sanitizer guidance
proves insufficient for bridging the gap
from sink reachability to successful exploitation.
This limited effectiveness stems from
the semantic complexity of exploitation conditions:
even with fine-grained feedback,
traditional fuzzing strategies struggle to synthesize inputs
that satisfy intricate sink-specific requirements.

To understand the root causes of exploitation failures,
we performed deeper analysis
on the reached-but-unexploited vulnerabilities.
The analysis identifies three primary categories:
\ding{172} \emph{Missing instrumentation} (1/21):
sinks used APIs not covered by value profile instrumentation,
leaving fuzzers without exploitation guidance.
\ding{173} \emph{Insufficient feedback depth} (2/21):
complex sanitizer conditions required longer input sequences
than value profile mechanisms could effectively guide.
\ding{174} \emph{Complex exploitation logic} (18/21):
this represents the most common failure mode.
Exploitation required reasoning about intricate conditions,
multiple API interactions,
or specific input formats
that traditional fuzzing could not synthesize
even with value profile feedback.
These vulnerabilities demand semantic understanding
of sink-specific exploitation requirements.
For instance,
constructing XML payloads that satisfy type constraints
for deserialization vulnerabilities,
or crafting paths that bypass sanitization
for path traversal attacks.

\PP{Implications}
This empirical study shows the improvement space for Java vulnerability detection and motivates us to build a system that distinctly addresses both the reachability and exploitability challenges.
With that understanding, we seek the opportunities to contextualize proper sink knowledge for both sub-problems with the help of LLMs' semantic-aware capabilities.

\begin{figure}[t]
  \centering
  \scriptsize
  \input{code/jenkins-doexec-commandutils.java}
  \vspace{0.5em}
  \RecustomVerbatimEnvironment{Verbatim}{Verbatim}{firstnumber=11}
  \input{code/jenkins-createutils.java}
  \caption{Command injection vulnerability in Jenkins
from AIxCC semifinal exemplar.
The vulnerability requires satisfying multiple conditions
to reach the \cc{ProcessBuilder} sink (line 20)
and specific input properties to trigger exploitation (line 23).}
  \label{fig:jenkins-example}
\end{figure}

\begin{figure*}
    {\centering
      \includegraphics[scale=0.67]{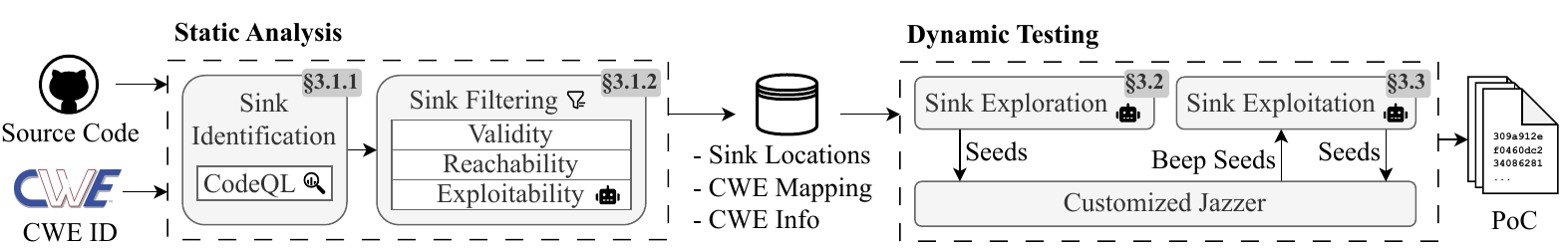}
    \par}

    \caption{\label{fig:overall-design} Overall design of \sys. Robot icons indicate LLM-based components.}
\end{figure*}

\subsection{A Motivating Example}

We illustrate the key sink knowledge
that \sys leverages through a concrete example
from the AIxCC competition's semifinal exemplars
(\autoref{fig:jenkins-example}).

\PP{Vulnerability Description}
The vulnerability implements a backdoor
enabling OS command injection through crafted HTTP requests.
The \cc{doExecCommandUtils} method (lines 1--10)
checks for an HTTP header \cc{x-evil-backdoor}
whose value must match the SHA-256 hash of \cc{"breakin the law"}.
When both conditions hold,
execution invokes \cc{createUtils},
which constructs a \cc{ProcessBuilder} with attacker-controlled arguments.
The \cc{ProcessBuilder} constructor at line 20 is the sink, \emph{i.e.}, a
security-sensitive API where attacker-controllable arguments
enable command execution.
Jazzer's sanitizer detects exploitation
when the command begins with \cc{"jazze"} (line 23).

\PP{Call Paths Encode Reachability Constraints}
Reaching the sink requires satisfying multiple constraints:
\ding{172} the HTTP request must contain header \cc{x-evil-backdoor} (line 5);
\ding{173} its SHA-256 hash must match \cc{"breakin the law"} (line 8);
and \ding{174} the command string must be non-empty (line 12).
These cryptographic and string comparison constraints
are difficult for coverage-guided fuzzing to satisfy through random mutation.
Our key finding is that
the call path from harness entry to sink
captures critical constraint information.
Here, the path through \cc{doExecCommandUtils} to \cc{createUtils}
reveals which methods must be traversed,
what input properties they examine,
and what conditions must hold.
This suggests a call path-based LLM agent
to solve reachability.
The agent collects information along the call path
and reasons about code structure,
identifying input format and critical conditions,
\emph{e.g.}, header name and hash comparison,
to synthesize satisfying inputs.

\PP{Beep Seeds Provide Exploitation Context}
Once the sink is reached,
exploitation requires the command string to begin with \cc{"jazze"}
(line 23).
Jazzer's value profile feedback solves this simple case.
However,
complex requirements expose the exploitability challenge.
For instance,
the exploitable input may not directly map to input bytes,
or additional constraints may exist.
The beep seed, \emph{i.e.}, the input that reaches the sink, provides
valuable context for exploitation.
It carries concrete input bytes,
execution stack trace,
and precise sink location.
Our key idea is to provide the LLM with debugging tools
and vulnerability-specific information,
enabling it to reason from the beep seed
and iteratively generate and debug exploit attempts.
This leverages dynamic information grounded in concrete program state
rather than requiring reasoning about all possible paths.
The beep seed thus offers an ideal starting point
for targeted exploit generation.

\subsection{Overview of \sys}
\label{s:prelim:overview}

Based on these findings,
we design and implement \sys,
a collaborative framework
that combines LLM-based agents with coverage-guided fuzzing
for vulnerability discovery.
The core design principle is to contextualize sink knowledge
from the target project
and use it to guide LLM reasoning
toward effective vulnerability detection.
Concretely,
\sys decomposes sink-based vulnerability discovery
into reachability and exploitability sub-tasks,
each handled by agents
grounded in task-specific context:
exploration agents use call-path code context
to solve reachability constraints,
while exploitation agents use beep-seed execution traces
to reason about exploitation conditions.
As shown in \autoref{fig:overall-design},
\sys realizes this by first detecting sinks (identification and filtering),
then deploying sink exploration and sink exploitation agents
concurrently with the fuzzer.

Given a target project and a specified CWE type,
\sys first scans and filters sinks with high exploitation potential.
This process begins by using CodeQL's sink database
to collect all candidate sink call sites.
\sys then applies multi-dimensional filtering
to eliminate irrelevant candidates.
This includes removing invalid sinks
with constant arguments or located in test code,
filtering unreachable sinks based on call graph analysis,
and assessing exploitability
through LLM-based contextual reasoning.

Once the sink list is determined,
\sys's exploration and exploitation agents
launch independently and run concurrently with the fuzzer instance.
The exploration agent analyzes call paths to sinks,
collects relevant path constraints and code context,
and iteratively generates inputs designed to reach target sinks.
All generated inputs are incorporated into the fuzzer's corpus.
The exploitation agent receives sink-reaching inputs,
\emph{i.e.}, beep seeds,
from the modified fuzzer runtime.
For each beep seed,
the agent analyzes the concrete execution context
and attempts to synthesize exploits
based on CWE-specific vulnerability knowledge.
All exploitation outputs are likewise shared with the fuzzer.
Throughout this process,
agents and non-LLM techniques cooperate as mutually beneficial:
agents refine CodeQL's static results
and contribute semantically grounded inputs,
while the fuzzer provides brute-force exploration
and beep seeds that ground the agents' reasoning.
When successful,
vulnerability discoveries manifest as sanitizer violations
detected by the fuzzer.

\section{Methodology}
\label{s:design}

This section details the design and implementation
of \sys's three core components introduced in \autoref{s:prelim:overview}.
We present sink detection (\autoref{s:design:detection}),
which identifies and filters candidate vulnerability locations;
sink exploration (\autoref{s:design:exploration}),
which generates inputs to reach identified sinks;
and sink exploitation (\autoref{s:design:exploitation}),
which develops proof-of-concept exploits based on sink-reaching inputs.

\subsection{Sink Detection}
\label{s:design:detection}

Given a target project and a user-specified CWE type,
the sink detection component identifies high-potential sinks
for subsequent dynamic analysis.
This task presents two key challenges in real-world scenarios.
First, we must establish a scalable approach to mapping CWE types to sink APIs,
one that accommodates both the extension of sink definitions within supported CWEs
and the addition of new CWE types.
Second, we must filter the identified sinks to reduce noise in real-world projects,
as the computational cost of downstream LLM-based analysis
scales linearly with the number of candidate sinks.

\subsubsection{CWE-Specific Sink API Call Site Extraction}

We address the first challenge by treating CodeQL as a comprehensive sink database.
CodeQL maintains thousands of sink definitions and patterns,
even including AI-generated sinks,
with ongoing community-driven updates and refinements.
Reusing this mature infrastructure
represents a more practical choice
than building a sink database from scratch
or employing large-scale LLM-based code scanning.

However, directly applying CodeQL's standard queries would undermine our objectives.
These queries implement conservative taint analysis
that reports sinks only when potentially malicious data flows from predefined sources,
\emph{e.g.}, network connections or user inputs.
While this design prioritizes precision in traditional static analysis contexts,
it introduces false negatives in our scenarios.
On one hand, attacker-controlled input originates from fuzzing harnesses may not align with CodeQL's predefined sources.
On the other hand, its inherent taint analysis may miss legitimate sinks due to indirect data flows or complex transformations.

We therefore treat CodeQL mainly as a sink database
and extract sink API call sites directly,
skipping any filtering logic it implements.
This requires rewriting CodeQL query scripts on a per-CWE basis.
The rewriting complexity depends on the query structure.
For queries where sink APIs are fully defined in YAML files
(CodeQL's data abstraction format)
and decoupled from the query logic,
we simply remove the taint-based filters to obtain all call sites.
For queries where sink APIs are hardcoded within the query logic,
we perform more fine-grained rewrites to extract the relevant patterns.
This approach maximizes the extensibility of query scripts within each CWE
while requiring only one-time engineering effort per CWE
in most cases.

\subsubsection{High-Potential Sink Identification}

The comprehensive sink extraction approach maximizes recall
but introduces substantial noise.
We address the second challenge
through multi-dimensional filtering
that eliminates invalid, unreachable, and unexploitable sinks
while preserving true vulnerabilities.
The filtering pipeline applies these filters sequentially,
terminating early when the number of remaining candidates falls below a manageable threshold
(10 in this manuscript, but configurable).
This early termination prevents unnecessary filtering
when the downstream LLM analysis cost is already affordable.

\PP{Invalid Sink Filtering}%
We first eliminate sinks
where security-sensitive API parameters receive only constant values.
Such invocations cannot be exploited through dynamic input manipulation,
as the argument values remain invariant across executions.
We leverage CodeQL's taint analysis infrastructure
to determine whether constant values flow to sensitive API parameters.
Additionally, we exclude sinks located within test code,
as these locations are not present in production deployments.
We employ a conservative heuristic:
filtering any sink whose class name contains \cc{Test}
or whose file path includes \cc{/test/}.
This filtering stage serves as a lightweight step using deterministic heuristics to filter out obvious false positives.

\PP{Unreachable Sink Filtering}%
We employ call graph analysis to identify unreachable sinks.
Sinks unreachable from the fuzzing harness entry point
cannot be exercised during fuzzing and can be safely eliminated.
We construct a call graph from the harness entry method
and retain only those sinks that appear as reachable nodes within this graph.
However, call graph construction in Java faces inherent limitations
due to dynamic language features such as reflection and dynamic class loading.
To mitigate the risk of incorrectly filtering true positives
due to call graph incompleteness,
we implement a conservative fallback strategy:
if reachability analysis eliminates all candidates,
we revert to the results from the previous filtering stage.

\PP{Unexploitable Sink Filtering}%
We employ an LLM-based agent to perform contextual exploitability assessment.
Determining whether a sink is unexploitable
represents a complex and challenging classification problem
that requires reasoning about code semantics and control flow.
We leverage LLMs to statically identify sinks
that can be confidently classified as non-exploitable
based on their surrounding code context.
The agent-based design enables flexible exploration of the codebase
to gather necessary context for making informed classification decisions,
rather than relying on fixed heuristics or analysis patterns.

The agent is designed with tools for autonomous codebase exploration,
including capabilities to read files, navigate directories, and search for code patterns.
The agent operates through a two-phase workflow.
In the code exploration phase,
it iteratively gathers relevant code context
by examining the immediate neighborhood of the sink,
analyzing data flow along the call path,
and inspecting the harness implementation.
It constructs a structured exploitability assessment report
documenting all collected evidence and reasoning.
In the decision phase,
the agent reviews this report and renders a binary classification decision.
A sink is classified as non-exploitable
only when it identifies concrete evidence,
such as constant-only data flows to sensitive parameters,
invariant checks that prevent malicious values,
or control flow constraints that prevent attacker-controlled data from reaching the sink.

\begin{algorithm}[t]
\caption{Sink exploration agent workflow}
\label{alg:sink-exploration}
\ForEach{sink $s$ in detected sinks}{
  $p \gets$ \textsc{SelectCallPath}($s$)\label{alg:line:path-select}\;
  $attempts \gets 0$, $feedback \gets \emptyset$\;
  \While{$attempts < maxIterations$}{
    $input \gets$ \textsc{GenerateInput}($p$, $feedback$)\label{alg:line:gen-loop}\;
    $reached \gets$ \textsc{ValidateInput}($input$, $s$)\label{alg:line:validate}\;
    \If{$reached$}{
      \textbf{break}\;
    }
    $feedback \gets$ \textsc{AnalyzeProgress}($input$, $p$)\label{alg:line:progress}\;
    $attempts \gets attempts + 1$\;
  }
}
\end{algorithm}

\subsection{Sink Exploration}
\label{s:design:exploration}

Coverage-guided fuzzing effectively explores reachable code
but is limited by complex constraints
such as cryptographic checks, input format requirements, or validation logic
that cannot be efficiently satisfied through random mutation.
The sink exploration component targets to address this challenge
by leveraging LLM-based reasoning to analyze call paths
and generate inputs designed to reach identified sinks.
The component runs concurrently with fuzzer,
contributing generated inputs to fuzzer's corpus as seeds for mutation-based fuzzing.

Our core insight is that
each call path from entry point to sink
naturally encodes the essential constraints for reaching that sink.
A call path is a sequence of function calls identified from the call graph.
While multiple execution paths may exist from entry to sink,
each call path captures the high-level structure of these executions, which is a compression of the underlying control flow.
By extracting code context along the call path,
an LLM agent can reason about input format requirements,
critical branching conditions,
and semantic constraints
to synthesize inputs likely to reach the target sink.

\autoref{alg:sink-exploration} presents our approach.
For each sink,
we first select a representative call path (\autoref{alg:line:path-select}),
then iteratively generate inputs guided by this path (\autoref{alg:line:gen-loop}).
Each generated input undergoes validation (\autoref{alg:line:validate})
to determine whether it successfully reaches the sink.
When validation fails,
we analyze reachability progress (\autoref{alg:line:progress})
to provide feedback on how far along the call path execution proceeded,
enabling the agent to refine subsequent attempts.
This iteration continues until either the sink is reached
or the maximum iteration budget is exhausted.

\PP{Call Path Selection}%
For each sink,
static analysis may identify one or more call paths from the harness entry point.
To manage computational costs when dealing with numerous sinks,
we select a single representative path per sink
rather than attempting input generation for all paths.
We apply a two-level ranking to prioritize paths.
First, paths with taint-based data flow evidence rank above call graph-only paths,
as taint analysis indicates that attacker-controlled input may influence the sink.
Second, within each category, we prioritize shorter paths,
as fewer intermediate function calls imply fewer branching conditions
and thus simpler constraint satisfaction problems.
The top-ranked path serves as the basis for subsequent input generation.

\PP{Input Generation}%
For the selected call path,
the LLM agent performs function-by-function code context collection.
The agent reads source code for each function along the path,
gathering information about input format requirements,
validation logic,
branching conditions,
etc.
The agent then synthesizes an input designed to satisfy the path constraints.
Rather than directly producing byte sequences,
we require the agent to generate a Python script that constructs the input.
The consideration behind this script-based approach is to leverage Python's expressiveness for complex input generation logic, and to serve as a more explainable base for iteration when validation fails.

\PP{Input Validation}%
To verify whether a generated input successfully reaches the target sink,
we employ debugger-based validation.
We execute the input using the Java debugger (JDB),
setting a breakpoint at the sink location.
If execution hits the breakpoint,
the input is confirmed to reach the sink
and is added to Jazzer's corpus as a beep seed candidate.
If execution does not reach the sink,
we proceed to reachability progress analysis.

\PP{Reachability Progress Analysis}%
When an input fails to reach the target sink,
we provide fine-grained feedback
to guide subsequent generation attempts.
The debugger tracks which functions along the call path were reached during execution,
identifying the deepest function node successfully entered
before execution diverged from the intended path.
This node-level reachability information is returned to the agent as feedback,
enabling it to understand which constraints were satisfied
and where execution deviated.
The agent takes this feedback in its next input generation attempt.

\begin{algorithm}[t]
\caption{Sink exploitation agent workflow}
\label{alg:sink-exploitation}
\While{fuzzer is running}{
  \textsc{UpdateBeepSeeds}()\label{alg:line:update-beeps}\;
  $beep \gets$ \textsc{ScheduleBeepSeed}()\label{alg:line:schedule-beeps}\;
  $attempts \gets 0$, $candidates \gets \emptyset$\;
  \While{$attempts < maxAttempts$}{\label{alg:line:gen-exploit-loop}
    $poc \gets$ \textsc{GenerateExploit}($beep$)\;
    $candidates \gets candidates \cup \{poc\}$\;
    \If{$poc$ triggers sanitizer}{
      \textbf{break}\;
    }
    $attempts \gets attempts + 1$\;
  }
  \textsc{NoCovFuzz}($beep$, $candidates$)\label{alg:line:no-cov-fuzz}\;
}
\end{algorithm}

\subsection{Sink Exploitation}
\label{s:design:exploitation}

Exploitation is a complex task
requiring deep understanding of vulnerability mechanisms
and precise manipulation of program state.
We address this challenge
by leveraging beep seeds to provide concrete context
that reduces the complexity of LLM-based exploit generation.
A beep seed provides specific input bytes that reach the sink,
dynamic execution details including stack traces and program state,
the expected CWE type,
and the target sanitizer conditions.
Rather than reasoning abstractly about all possible exploitation scenarios,
we construct an agent that works from this concrete foundation
to iteratively develop exploits.
We create a dynamic interactive environment
that enables the agent to experiment and refine attempts,
similar to how human security researchers use debugging tools
to develop exploits from initial sink-reaching inputs.

\autoref{alg:sink-exploitation} presents the exploitation workflow.
The agent operates in a continuous loop,
receiving beep seeds from the fuzzer (\autoref{alg:line:update-beeps}),
selecting promising candidates for exploitation attempts (\autoref{alg:line:schedule-beeps}),
and generating proof-of-concept exploits through iterative refinement (\autoref{alg:line:gen-exploit-loop}).
All generated candidates,
whether successful or unsuccessful,
feed into a specialized fuzzing stage (\autoref{alg:line:no-cov-fuzz})
which aims to refine almost-working exploits into successful ones.
In the end, all candidates will always be synced back to the fuzzer's corpus for further mutation-based fuzzing.

\PP{Beep Seed Collection}%
The agent monitors beep seeds produced by the fuzzing component.
We implement this through fuzzer customization.
Specifically, we instrument all identified sinks in the target program,
capturing execution details when fuzzing reaches these locations.
For each distinct sink-reaching input,
the fuzzer saves the input bytes,
stack trace,
sink location,
and associated CWE type.
These beep seed packages are continuously streamed to the exploitation agent
throughout the fuzzing campaign.

\PP{Beep Seed Scheduling}%
The agent groups collected beep seeds by stack trace,
treating different execution paths to the same sink as distinct groups.
Scheduling prioritizes groups
that have been attempted fewer times,
subject to a maximum schedule limit per group
(set to 1 in our experiments).
Within each selected group,
the agent randomly chooses the beep seed for the next exploitation attempt.
This strategy ensures diverse coverage of different sink-reaching execution paths
while preventing excessive resource expenditure on any single path.

\PP{Exploit Generation}%
For the selected beep seed,
the agent receives the input file location,
stack trace,
and basic sink information.
The agent uses file exploration tools
to gather additional code context relevant to exploitation,
following the stack trace to collect code surrounding each frame.
Similar to the exploration agent,
the agent generates a Python script
that constructs the exploit input.
The script is executed to test whether it triggers the sanitizer.
If unsuccessful,
the execution log from the most recent attempt
is incorporated into the context for the next generation iteration.
This feedback loop enables the agent to learn from failed attempts
and refine its understanding of the exploitation constraints.

\PP{No-Coverage Fuzzing}%
After the generation loop completes or reaches the attempt limit,
we execute a specialized fuzzing phase
using both the original beep seed and all generated candidates as the corpus.
This phase addresses cases where LLM-generated answers approximate
but do not fully satisfy the exploit conditions.
Traditional mutation-based fuzzing can bridge this gap.
We disable coverage feedback in this stage
because we do not seek exploring new coverage (note that the sink is already reached).
Instead, we retain value profile feedback,
which provides fine-grained guidance toward triggering the sanitizer.
This focused fuzzing runs for a short duration
to refine almost-working exploits into successful ones.
Eventually, all generated candidates are synced back to the fuzzer's corpus for further mutation-based fuzzing.

\section{Evaluation}
\label{s:eval}

\PP{Implementation}%
We implement \sys as a fully dockerized system
comprising \numloclines lines of code
across sink detection (341\,LoC CodeQL, 295\,LoC Python),
sink exploration (10,715\,LoC Python, 430\,LoC Java),
and sink exploitation (2,481\,LoC Python, plus Jazzer modifications in Java).
The system uses LangChain~\cite{langchain} and LangGraph~\cite{langgraph} for agent development,
CodeQL in sink detection,
and Joern~\cite{joern} in sink exploration
(CHA~\cite{cha} and RTA~\cite{rta} enabled in call graph construction).
\sys is fully compatible with OSS-Fuzz projects.

For LLM interactions,
we employ LiteLLM as a unified client interface
with automatic retry mechanisms
and client-side cost tracking.
Each agent operates with a configurable iteration limit.
\sys currently supports \numcwes CWE types (\autoref{t:cwes});
additional CWE coverage requires
adding its CodeQL query script
and vulnerability description.
Per-CWE effort is minimal:
most sink extraction queries are approximately five lines of CodeQL
that reuse the existing CWE-sink database
while bypassing default taint filtering.
The JDB-based input validation (\autoref{s:design})
adds little overhead,
with 90th/99th-percentile execution times of 6 and 19 seconds;
\sys limits JDB usage to breakpoint-based reachability probing
to minimize this cost.


\PP{Research Questions}
\begin{itemize}
    \item \textbf{RQ1:} How does \sys perform compared to the state of the art?
    \item \textbf{RQ2:} How does each component contribute to \sys's overall effectiveness?
    \item \textbf{RQ3:} How effective are the LLM agents within each component?
    \item \textbf{RQ4:} How is \sys perceived by industry practitioners?
\end{itemize}

\begin{table}[t]
\centering
\caption{CWE types currently supported by \sys.}
\label{t:cwes}
\begin{threeparttable}
\scriptsize
\setlength{\tabcolsep}{3pt}
\begin{tabular}{l@{\ \ }p{5cm}}
\toprule
\textbf{CWE-ID} & \textbf{Short Name} \\
\midrule
CWE-022 & Path Traversal \\
\rowcolor{gray!10}
CWE-078 & OS Command Injection \\
CWE-089 & SQL Injection \\
\rowcolor{gray!10}
CWE-090 & LDAP Injection \\
CWE-094 & Code Injection \\
\rowcolor{gray!10}
CWE-117 & Log Injection \\
CWE-470 & Unsafe Reflection \\
\rowcolor{gray!10}
CWE-502 & Deserialization of Untrusted Data \\
CWE-611 & Improper Restriction of XML External Entity Reference (XXE) \\
\rowcolor{gray!10}
CWE-643 & XPath Injection \\
CWE-730 & Denial of Service (Regular Expression Injection) \\
\rowcolor{gray!10}
CWE-918 & Server-Side Request Forgery \\
\bottomrule
\end{tabular}
\end{threeparttable}

\end{table}

\PP{Benchmark}%
No existing benchmark supports
dynamic sink-based vulnerability discovery:
JQF~\cite{jqf} covers only uncaught exceptions,
and Iris~\cite{iris} targets static analysis
with 4~CWE types and no harnesses or proof-of-concept exploits.
We therefore constructed a new dataset
to evaluate \sys on realistic vulnerability scenarios.
Five security researchers
(5--10 years experience each)
contributed 15 person-weeks
to develop the benchmark
using well-established open-source Java projects.

The resulting benchmark comprises \numcpvs vulnerabilities
across \numprojects projects,
spanning \numcwes CWE types
and reachable from \numharnesses OSS-Fuzz-compatible harnesses
(\autoref{t:dataset}),
providing broad coverage
for evaluating generality across vulnerability classes.
Researchers populated the dataset through three approaches:
\numcpvsCVEs CVE-based vulnerabilities
with custom-built harnesses covering \numcpvsCVEsCWEs CWE types,
\numcpvsAIxCC vulnerabilities
from DARPA AIxCC Competition~\cite{aixcc} challenges
(reused with organizer approval),
and \numcpvsSynthetic manually injected synthetic vulnerabilities
modeled after real-world patterns.
Jenkins' plugin architecture serves as a natural base
to inject multiple vulnerabilities within a single harness.
All harnesses invoke high-level API functions
in conformant ways
while allowing underlying vulnerabilities to be triggered.
The dataset includes ground truth annotations
for vulnerability locations, types, stack traces,
and proof-of-concept exploits.

\begin{table}[t]
\centering
\caption{Dataset overview
showing the projects evaluated with \sys,
including lines of code,
number of harnesses,
number of vulnerabilities,
and expected CWE types.}
\label{t:dataset}
\begin{threeparttable}
\scriptsize
\setlength{\tabcolsep}{3pt}
\begin{tabular}{l@{\ }r@{\ \ }r@{\ \ }r@{\ \ }p{4.1cm}}
\toprule
\textbf{Project} & \textbf{LoC} & \textbf{\#Har} & \textbf{\#V} & \textbf{Expected CWEs} \\
\midrule
activemq & 807 & 2 & 1 & CWE-470 \\
\rowcolor{gray!10}
apache-cc & 85 & 4 & 4 & CWE-022, CWE-502, CWE-078, CWE-918 \\
batik & 295 & 2 & 1 & CWE-918 \\
\rowcolor{gray!10}
bcel & 55 & 2 & 1 & CWE-022 \\
cxf & 1,111 & 1 & 1 & CWE-918 \\
\rowcolor{gray!10}
feign & 54 & 1 & 1 & CWE-730 \\
fuzzy & 2 & 2 & 1 & CWE-643 \\
\rowcolor{gray!10}
geonetwork & 1,049 & 2 & 1 & CWE-078 \\
imaging & 48 & 4 & 3 & CWE-078, CWE-643, CWE-730 \\
\rowcolor{gray!10}
jackson-databind & 164 & 2 & 1 & CWE-470 \\
jakarta-mail-api & 75 & 2 & 1 & CWE-611 \\
\rowcolor{gray!10}
jenkins & 503 & 7 & 14 & CWE-022, CWE-078, CWE-089, CWE-090, CWE-094, CWE-117, CWE-470, CWE-502, CWE-643, CWE-730, CWE-918 \\
kylin & 394 & 1 & 1 & CWE-078 \\
\rowcolor{gray!10}
oripa & 999 & 2 & 1 & CWE-611 \\
pac4j & 57 & 1 & 1 & CWE-502 \\
\rowcolor{gray!10}
rdf4j & 113 & 1 & 3 & CWE-022, CWE-611 \\
shiro & 37 & 1 & 1 & CWE-502 \\
\rowcolor{gray!10}
tika & 261 & 10 & 10 & CWE-022, CWE-078, CWE-502, CWE-611, CWE-918 \\
widoco & 46 & 1 & 1 & CWE-022 \\
\rowcolor{gray!10}
xstream & 80 & 1 & 2 & CWE-470, CWE-611 \\
zookeeper & 190 & 3 & 3 & CWE-022, CWE-502, CWE-078 \\
\rowcolor{gray!10}
ztzip & 7 & 1 & 1 & CWE-022 \\
\midrule
\textbf{TOTAL} &  & \textbf{53} & \textbf{54} &  \\
\bottomrule
\end{tabular}
\begin{tablenotes}
\setlength{\itemindent}{-10pt}
\item \textbf{LoC}: Thousands of lines of code, \textbf{\#Har}: Number of harnesses, \textbf{\#V}: Number of vulnerabilities
\end{tablenotes}
\end{threeparttable}

\end{table}

\PP{Evaluation Setup}
We conduct all experiments
on two dedicated servers:
a dual AMD EPYC 9354 system (128 logical cores, 768\GB RAM)
and a single AMD EPYC 7543 system (64 logical cores, 256\GB RAM),
both running Ubuntu 24.04.
Each harness runs three fuzzing instances (three cores) for 12 hours,
with all configurations of each challenge
executed on the same server
to ensure consistent CPU specifications.
%
%
We configure Jazzer to maximize vulnerability detection:
25-second timeout detection threshold,
persistent execution after crash discovery,
value profile instrumentation,
8\GB maximum RSS,
corpus reloading every 30 seconds,
and 1\MB input size limit.
All remaining Jazzer settings,
including default seeds and dictionaries,
follow OSS-Fuzz defaults
unless overridden by AIxCC challenge specifications.

We evaluate \sys with LLM models from major providers
including open source models
and conduct ablation studies with each key component disabled.
LLM agents operate with default temperature settings,
token limits matching context size,
and a maximum of 30 loop iterations
with automatic retry on transient failures.
Static analysis employs a threshold of 10 candidate sinks;
the exploitation kit allocates 5 minutes
of coverage-feedback-free fuzzing
for each generated seed.

\PP{LLM Models}%
We evaluate \sys with seven LLMs
spanning three flagship models
(GPT-5, Gemini~2.5~Pro, Claude Sonnet~4.5),
two lightweight models
(GPT-5-nano, Gemini~2.5~Flash~Lite),
and two open-weight models
(GPT-OSS-120B, GLM-5).
\autoref{t:models} lists each model's provider
and training data cutoff date;
all cutoffs predate the release of our benchmark.
We discuss potential model contamination
in \autoref{s:discussion}.

\begin{table}[t]
\centering
\caption{LLM models evaluated with \sys.}
\label{t:models}
\scriptsize
\begin{threeparttable}
\begin{tabular}{llll}
\toprule
\textbf{Model} & \textbf{Provider} & \textbf{Tier} & \textbf{Cutoff} \\
\midrule
GPT-5 & OpenAI & Flagship & Sep 2024 \\
\rowcolor{gray!10}
Gemini 2.5 Pro & Google & Flagship & Jan 2025 \\
Sonnet 4.5 & Anthropic & Flagship & Jul 2025 \\
\rowcolor{gray!10}
GPT-5-nano & OpenAI & Lightweight & May 2024 \\
Gemini 2.5 Flash Lite & Google & Lightweight & Jan 2025 \\
\rowcolor{gray!10}
GPT-OSS-120B & OpenAI & Open-weight & Jun 2024 \\
GLM-5 & Zhipu AI & Open-weight & N/A\tnote{$\dagger$} \\
\bottomrule
\end{tabular}
\begin{tablenotes}
\setlength{\itemindent}{-5pt}
\item[$\dagger$] Not disclosed; model released Feb 2026
\end{tablenotes}
\end{threeparttable}
\end{table}

\subsection{\sys Internals}

\autoref{t:internals} presents performance metrics
for \sys's three main components
across all \numprojects projects
when using GPT-5.
We analyze each stage's effectiveness
in filtering false positives,
reaching sinks, and exploiting sinks.

\PP{Sink Detection}%
Our CodeQL queries successfully identify all \numcpvs vulnerabilities
across the \numprojects projects
(\textbf{CQL Sinks, \#S}, \autoref{t:internals}),
demonstrating comprehensive coverage.
From an initial 8,262 potential sinks
(1:152 false positive ratio),
the three filtering stages
eliminate 95\% of false positives,
yielding 383 actionable sinks
(\textbf{Final, \#S}, \autoref{t:internals}).
Filtering retains 52 out of 54 expected vulnerabilities;
two false negatives stem from Joern's inability
to resolve reflective calls and lambdas,
a general limitation of Java static analysis.
LLM cost scales with analyzed sinks,
with command injection most expensive
due to iterative payload refinement
for Jazzer's sanitizer constraints.

\begin{figure}[t]
\centering
\footnotesize
\input{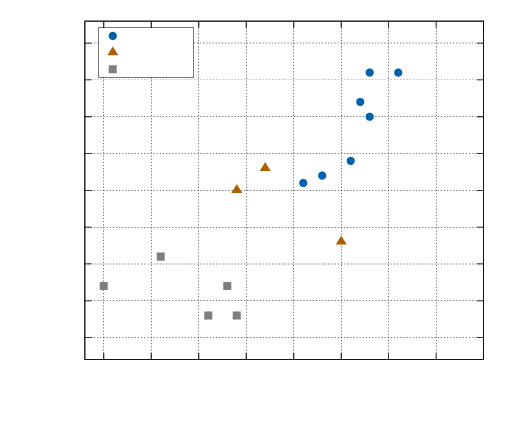}
\caption{Coordinate diagram
showing the relationship between vulnerabilities reached and exploited
for different tools and configurations.}
\label{f:tool-coords}
\end{figure}

\begin{table*}[t]
\centering
\caption{Internals of \sys
showing static analysis filtering stages
and dynamic analysis details
across \numprojects projects (\sysGPT).}
\label{t:internals}
\begin{threeparttable}
\scriptsize
\setlength{\tabcolsep}{2pt}
\begin{tabular}{l >{\raggedleft\arraybackslash}p{1.5em} || >{\raggedleft\arraybackslash}p{2em}>{\raggedleft\arraybackslash}p{2.5em} | >{\raggedleft\arraybackslash}p{2em} | >{\raggedleft\arraybackslash}p{2em} | >{\raggedleft\arraybackslash}p{2em} | >{\raggedleft\arraybackslash}p{2em}>{\raggedleft\arraybackslash}p{2.5em} | >{\raggedleft\arraybackslash}p{3.5em} || >{\raggedleft\arraybackslash}p{2.5em}>{\raggedleft\arraybackslash}p{2.5em} | >{\raggedleft\arraybackslash}p{2.5em} | >{\raggedleft\arraybackslash}p{2.0em}>{\raggedleft\arraybackslash}p{2.0em} | >{\raggedleft\arraybackslash}p{4.0em} || >{\raggedleft\arraybackslash}p{2.5em}>{\raggedleft\arraybackslash}p{2.5em}>{\raggedleft\arraybackslash}p{2.5em} | >{\raggedleft\arraybackslash}p{2.0em}>{\raggedleft\arraybackslash}p{2.0em} | >{\raggedleft\arraybackslash}p{3.5em}}
\toprule
 &  & \multicolumn{8}{c}{\textbf{Sink Detection}} & \multicolumn{6}{c}{\textbf{Sink Exploration}} & \multicolumn{6}{c}{\textbf{Sink Exploitation}} \\
\cmidrule(lr){3-10} \cmidrule(lr){11-16} \cmidrule(lr){17-22}
 &  & \multicolumn{2}{c}{\textbf{CQL Sinks}} & \multicolumn{3}{c}{\textbf{Filtered Sinks}} & \multicolumn{2}{c}{\textbf{Final}} &  & \multicolumn{2}{c}{\textbf{Paths}} & \multicolumn{1}{c}{\textbf{Seeds}} & \multicolumn{2}{c}{\textbf{Result}} & \multicolumn{1}{c}{} & \multicolumn{3}{c}{\textbf{Seeds}} & \multicolumn{2}{c}{\textbf{Result}} & \multicolumn{1}{c}{} \\
\cmidrule(lr){3-4} \cmidrule(lr){5-7} \cmidrule(lr){8-9} \cmidrule(lr){11-12} \cmidrule(lr){13-13} \cmidrule(lr){14-15} \cmidrule(lr){17-19} \cmidrule(lr){20-21}
\textbf{Project} & \textbf{\#V} & \textbf{\#V} & \textbf{\#S} & \textbf{Inv} & \textbf{UR} & \textbf{UE} & \textbf{\#V} & \textbf{\#S} & \textbf{LLM \$} & \textbf{Taint} & \textbf{CG} & \textbf{Gen} & \textbf{R} & \textbf{E} & \textbf{LLM \$} & \textbf{Ana} & \textbf{Grp} & \textbf{Gen} & \textbf{E\textsubscript{LLM}} & \textbf{E} & \textbf{LLM \$} \\
\midrule
activemq & 1 & 1 & 131 & 12 & 82 & 23 & 1 & 14 & 6.29 & 4 & 24 & 26 & 1 & 0 & 100.88 & 55 & 48 & 347 & 1 & 1 & 108.82 \\
\rowcolor{gray!10}
apache-cc & 4 & 4 & 51 & 6 & 33 & 0 & 4 & 12 & 0.00 & 2 & 10 & 14 & 3 & 0 & 19.21 & 15 & 3 & 102 & 3 & 3 & 21.06 \\
batik & 1 & 1 & 59 & 30 & 0 & 0 & 1 & 29 & 9.33 & 0 & 0 & 0 & 0 & 0 & 0.00 & 71 & 2 & 209 & 0 & 1 & 116.68 \\
\rowcolor{gray!10}
bcel & 1 & 1 & 61 & 29 & 23 & 0 & 1 & 9 & 0.00 & 0 & 14 & 17 & 1 & 0 & 25.55 & 37 & 1 & 190 & 1 & 1 & 54.17 \\
cxf & 1 & 1 & 197 & 6 & 120 & 42 & 1 & 29 & 32.70 & 1 & 25 & 41 & 1 & 0 & 221.67 & 2 & 1 & 4 & 1 & 1 & 0.10 \\
\rowcolor{gray!10}
feign & 1 & 1 & 25 & 23 & 0 & 0 & 1 & 2 & 0.00 & 0 & 1 & 0 & 0 & 0 & 0.21 & 0 & 0 & 0 & 0 & 0 & 0.00 \\
fuzzy & 1 & 1 & 1 & 0 & 0 & 0 & 1 & 1 & 0.00 & 0 & 2 & 1 & 1 & 0 & 0.47 & 4 & 2 & 15 & 1 & 1 & 2.90 \\
\rowcolor{gray!10}
geonetwork & 1 & 1 & 1 & 0 & 0 & 0 & 1 & 1 & 0.00 & 0 & 2 & 3 & 0 & 0 & 1.22 & 0 & 0 & 0 & 0 & 0 & 0.00 \\
imaging & 3 & 3 & 17 & 13 & 0 & 0 & 3 & 4 & 0.00 & 0 & 0 & 0 & 0 & 0 & 0.00 & 0 & 0 & 0 & 0 & 0 & 0.00 \\
\rowcolor{gray!10}
jackson-databind & 1 & 1 & 27 & 1 & 6 & 0 & 1 & 20 & 8.57 & 0 & 40 & 49 & 0 & 0 & 206.42 & 31 & 2 & 81 & 1 & 1 & 56.99 \\
jakarta-mail-api & 1 & 1 & 9 & 0 & 0 & 0 & 1 & 9 & 0.00 & 0 & 0 & 0 & 0 & 0 & 0.00 & 0 & 0 & 0 & 0 & 0 & 0.00 \\
\rowcolor{gray!10}
jenkins & 14 & 14 & 4428 & 2655 & 1601 & 105 & 13 & 65 & 33.00 & 24 & 27 & 56 & 10 & 0 & 107.97 & 90 & 24 & 928 & 10 & 10 & 107.07 \\
kylin & 1 & 1 & 148 & 145 & 0 & 0 & 1 & 3 & 0.00 & 0 & 3 & 3 & 1 & 0 & 2.03 & 7 & 1 & 36 & 1 & 1 & 4.49 \\
\rowcolor{gray!10}
oripa & 1 & 1 & 2 & 0 & 0 & 0 & 1 & 2 & 0.00 & 0 & 2 & 1 & 1 & 0 & 0.38 & 0 & 0 & 0 & 0 & 0 & 0.00 \\
pac4j & 1 & 1 & 1 & 0 & 0 & 0 & 1 & 1 & 0.00 & 1 & 0 & 2 & 0 & 0 & 0.48 & 9 & 3 & 17 & 0 & 0 & 10.45 \\
\rowcolor{gray!10}
rdf4j & 3 & 3 & 78 & 1 & 51 & 4 & 3 & 22 & 15.38 & 2 & 16 & 29 & 3 & 0 & 70.94 & 21 & 3 & 286 & 3 & 3 & 18.09 \\
shiro & 1 & 1 & 9 & 1 & 0 & 0 & 1 & 8 & 0.00 & 0 & 0 & 0 & 0 & 0 & 0.00 & 0 & 0 & 0 & 0 & 0 & 0.00 \\
\rowcolor{gray!10}
tika & 10 & 10 & 2178 & 1427 & 245 & 319 & 9 & 111 & 234.22 & 0 & 104 & 236 & 7 & 0 & 534.09 & 84 & 13 & 454 & 4 & 4 & 59.50 \\
widoco & 1 & 1 & 62 & 18 & 40 & 0 & 1 & 4 & 0.00 & 2 & 2 & 4 & 1 & 0 & 2.29 & 10 & 1 & 113 & 1 & 1 & 8.27 \\
\rowcolor{gray!10}
xstream & 2 & 2 & 53 & 21 & 8 & 4 & 2 & 20 & 7.08 & 3 & 14 & 11 & 1 & 1 & 36.11 & 19 & 3 & 166 & 1 & 1 & 18.22 \\
zookeeper & 3 & 3 & 459 & 356 & 97 & 0 & 3 & 6 & 0.00 & 3 & 3 & 4 & 1 & 0 & 4.66 & 14 & 6 & 62 & 1 & 2 & 16.88 \\
\rowcolor{gray!10}
ztzip & 1 & 1 & 265 & 148 & 103 & 3 & 1 & 11 & 6.81 & 0 & 11 & 21 & 1 & 0 & 34.67 & 4 & 1 & 27 & 1 & 1 & 4.56 \\
\midrule
\textbf{TOTAL} & \textbf{54} & \textbf{54} & \textbf{8262} & \textbf{4892} & \textbf{2409} & \textbf{500} & \textbf{52} & \textbf{383} & \textbf{353.39} & \textbf{42} & \textbf{300} & \textbf{518} & \textbf{33} & \textbf{1} & \textbf{1369.24} & \textbf{473} & \textbf{114} & \textbf{3037} & \textbf{30} & \textbf{32} & \textbf{608.28} \\
\bottomrule
\end{tabular}
\begin{tablenotes}
\setlength{\itemindent}{-10pt}
\item \textbf{CQL Sinks}: Results of codeql sinks detection, \textbf{Filtered Out}: Sinks filtered out by our criteria, \textbf{Final}: Our results after filtering
\item \textbf{\#V}: Number of vulnerabilities, \textbf{\#S}: Number of sinks, \textbf{Inv}: Invalid sink filtering, \textbf{UR}: Unreachable sink filtering, \textbf{UE}: Unexploitable sink filtering
\item \textbf{Taint}: Taint tracking paths discovered, \textbf{CG}: Call graph paths discovered, \textbf{Gen}: Generated seeds, \textbf{R}: Reached sinks, \textbf{E}: Exploited  sinks
\item \textbf{Ana}: Analyzed seeds, \textbf{Grp}: Number of groups of seeds, \textbf{E\textsubscript{LLM}}: Sinks exploited by agent without fuzzer, \textbf{LLM \$}: Total LLM cost in USD
\end{tablenotes}
\end{threeparttable}

\end{table*}

\begin{figure*}[t]
\centering
\input{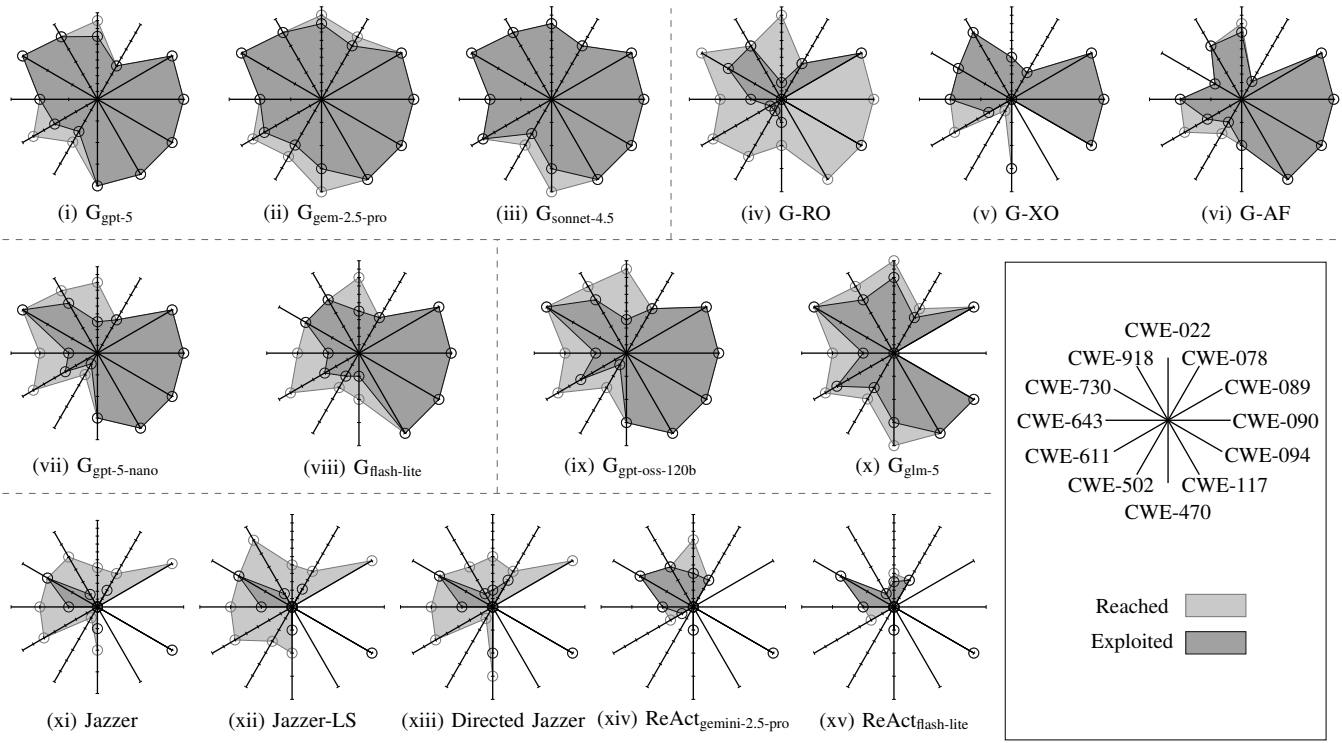}
\caption{Spider chart showing normalized reached and exploited vulnerabilities per CWE type for 
    baseline settings and different configurations of \sys.
    The value at each axis is calculated as the number of vulnerabilities reached/exploited
    divided by the total number of vulnerabilities
    for that CWE type.
    }
\label{f:spider-per-config}
\end{figure*}

\PP{Sink Exploration}%
The exploration agent
successfully reaches 33 out of 52 detected vulnerabilities
(\textbf{Result, R}, \autoref{t:internals}),
guiding fuzzing towards vulnerable sinks
without prior seed inputs.
Static taint analysis
fails to find paths for 13 projects
(\textbf{Paths, Taint} in \autoref{t:internals});
call graph fallback proves crucial,
enabling 13 additional vulnerabilities to be reached.
Six unreached vulnerabilities
stem from Joern's call graph limitations
with reflective calls,
not from \sys's approach itself.
The agent generates only 1.4 seeds per analyzed sink
(518 seeds for 383 sinks),
demonstrating the effectiveness
of path-aware, high-quality seed generation.
LLM costs total 1,369.24\USD (GPT-5),
suggesting future optimization
through improved path prioritization.

\PP{Sink Exploitation}%
The exploitation agent
generates working exploits
for 32 out of 42 reached vulnerabilities
(\textbf{Result, E}, \autoref{t:internals}),
achieving 76\% success.
The 32 successful exploits
are generated from 473 analyzed seeds,
after grouping them into 114 distinct contexts
based on their stack trace (\autoref{s:design:exploitation}).
LLM costs total 608.28\USD (GPT-5),
averaging 19.01\USD per exploited vulnerability
and 1.29\USD per analyzed seed.
Seed generation is inexpensive (0.20\USD per seed);
the majority of cost
stems from analyzing seeds
to craft exploit attempts.

\subsection{RQ1: Comparison with State of the Art}

\PP{Baselines}%
We compare \sys under different LLM configurations
against two baseline settings of Jazzer (see Evaluation Setup),
a modified version of Jazzer to support directed fuzzing,
and a simple ReAct agent.
The baseline configurations are:

\squishlist
    \item \textbf{\baseline}: Off-the-shelf Jazzer
    \item \textbf{\baselineLS}: Off-the-shelf Jazzer
    under large resource allocation (50 cores, 24 hours)
    \item \textbf{\directedJazzer}: Jazzer
    with directed fuzzing modification (AFLGO distance scheduling),
    manually configured ground truth sinkpoints
    \item \textbf{\reactPro / \reactFL}: ReAct LLM agent
    generating inputs given a sink location (no fuzzer);
    agent inputs are project information, vulnerable sink location, and CWE information;
    agent has access to source code and harness executor tool
\squishend

\noindent
We denote our tool as follows:

\squishlist
    \item \textbf{\sys\textsubscript{\textit{model}}}: \sys
    using the given LLM (\autoref{t:models})
\squishend

\noindent
For sink filtering,
we additionally compare against two static analysis tools:

\squishlist
    \item \textbf{CodeQL}: GitHub's semantic code analysis engine,
    using its built-in CWE-specific vulnerability queries
    \item \textbf{SpotBugs}: A widely used static bug finder for Java
\squishend

\begin{figure*}[t]
\centering
\includegraphics[width=\textwidth]{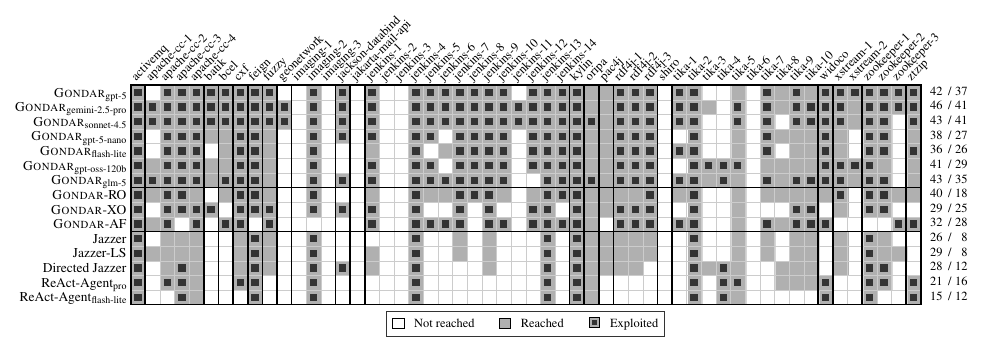}
\caption{Per-vulnerability matrix showing reached (light) and exploited (dark) status
for each of the 54 vulnerabilities across all configurations.}
\label{f:cpv-matrix}
\end{figure*}

\PP{Metrics and Visualization}
Since the ultimate goal is the discovery of vulnerabilities,
the number of vulnerabilities reached
and exploited
serve as our metrics of effectiveness.
We present overall results
in a coordinate diagram
(\autoref{f:tool-coords}),
arranging the different configurations
based on reached (x-axis)
and exploited (y-axis) vulnerabilities.
In addition, we provide per-project details
in \autoref{f:cpv-matrix}
and CWE-based breakdowns using
\autoref{f:spider-per-config}, (i)-(iii) and (vii)-(xv).

\PP{Overall Performance}
In terms of vulnerability detection,
\autoref{f:tool-coords} shows that \sys using any LLM
significantly outperforms the baseline Jazzer
by exploiting at least three times as many vulnerabilities (26 vs.\ 8).
\sysGemini
achieves the best performance,
reaching 46 vulnerabilities
and exploiting 41,
surpassing the baseline
by 20 reached
and 33 exploited vulnerabilities,
an over 4\x improvement in exploitation.
Even the least effective configuration
(\sysFlashLite)
significantly outperforms the baseline,
reaching 38.5\% more vulnerabilities
and exploiting 3.25 times as many.
Using flagship models, \sys reaches more vulnerabilities
in at least 9 out of 12 CWE categories
and exploits more vulnerabilities
in every CWE category.
These results demonstrate
the clear effectiveness of \sys's approach
in enhancing Java vulnerability discovery
over a wide range of CWE types
compared to state-of-the-art fuzzing.

On the other hand, \autoref{f:tool-coords} also
reveals that resource scaling does not solve the fundamental limitations of Java fuzzing.
The large-scale baseline
reaches 29 vulnerabilities
and exploits 8,
only marginally improving
over the standard baseline
(3 more reached, same exploited).
This plateau demonstrates that Jazzer rapidly saturates its search space regardless of computational resources, indicating an inherent constraint in its exploration strategy.
In our experiments,
50\% of challenges reach 95\% of their maximum coverage
within 15 minutes,
and 70\% within 6 hours.
These findings support our approach using qualitatively distinct techniques rather than quantitative resource scaling.
Directed Jazzer (reaches 27, exploits 12)
only slightly improves over plain Jazzer,
demonstrating that distance-based scheduling alone
is insufficient without LLM-assisted reasoning.
The ReAct agents
(\reactPro exploits 16; \reactFL exploits 12)
perform comparably to Directed Jazzer (exploits 12)
but far below \sys,
confirming that problem decomposition
and grounded reasoning,
not raw LLM capability,
drive \sys's effectiveness.

\PP{Analysis by LLM}
Analyzing \sys's performance
across different LLM models
reveals slight variations
in effectiveness.
Cross-vendor differences among flagship models
are minor (0--4 vulnerabilities),
whereas the gap between flagship and lightweight models
is larger (10--15 vulnerabilities),
suggesting that \sys's design
naturally benefits from stronger LLM reasoning.
While \sysGemini
achieves the best overall performance,
\sysSonnet
demonstrates the highest exploitation rate,
successfully exploiting all but two reached vulnerabilities.
Interestingly,
\sysSonnet
discovers one vulnerability
missed by \sysGemini;
this suggests that
no single model
is universally optimal,
and combining multiple models
may further enhance effectiveness in future work.

The reason for the performance differences
across models
appears to stem
from their varying strengths
and weaknesses
on different CWE types.
For instance,
\autoref{f:spider-per-config}
shows that
\sysGPT
has difficulties with CWE-078 (Command Injection),
making it the least effective
on that category, and overall.
Another challenging category
is CWE-502 (Deserialization of Untrusted Data),
which all models have difficulty with.
This is expected,
as exploiting such vulnerabilities
requires generating complex binary objects
as inputs,
which must also trigger Jazzer's deserialization sanitizer.
Our findings suggest that models
may have difficulties reaching 
and exploiting vulnerabilities of certain CWE types,
with \sysGemini being the overall most robust.

When analyzing the cost of different models,
we find that effectiveness scales with cost in our experiments:
\sysGemini
incurs the highest overall costs
(3,051.44\USD)
while achieving the best performance,
followed by \sysSonnet
(2,919.41\USD)
and \sysGPT
(2,428.83\USD).
This indicates
that investing in more capable models
yields better returns
in vulnerability discovery.

\begin{table}[t]
\centering
\caption{Cost analysis of different \sys configurations.}
\label{t:costs}
\resizebox{\columnwidth}{!}{
\begin{threeparttable}
\scriptsize
\setlength{\tabcolsep}{3pt}
\begin{tabular}{l|rrrr|r}
\toprule
 & \textbf{Sink} & \textbf{Sink} & \textbf{Sink} &  &  \\
 & \textbf{Filtering} & \textbf{Exploration} & \textbf{Exploitation} & \textbf{Fuzzing} & \textbf{Total} \\
\midrule
\sysGPT & \$353.39 & \$1,369.24 & \$608.28 & \$97.92 & \$2,428.83 \\
\rowcolor{gray!10}
\sysGemini & \$123.74 & \$2,530.98 & \$298.80 & \$97.92 & \$3,051.44 \\
\sysSonnet & \$751.93 & \$804.23 & \$1,265.33 & \$97.92 & \$2,919.41 \\
\rowcolor{gray!10}
\sysNano & \$4.60 & \$73.19 & \$6.23 & \$97.92 & \$181.93 \\
\sysFlashLite & \$40.99 & \$61.83 & \$6.86 & \$97.92 & \$207.60 \\
\rowcolor{gray!10}
\sysGPTOSS & \$1.74 & \$47.44 & \$1.13 & \$97.92 & \$148.23 \\
\sysGLM & \$19.32 & \$266.17 & \$8.90 & \$97.92 & \$392.31 \\
\rowcolor{gray!10}
\baseline & - & - & - & \$97.92 & \$97.92 \\
\baselineLS & - & - & - & \$3,263.95 & \$3,263.95 \\
\rowcolor{gray!10}
\directedJazzer & - & - & - & \$97.92 & \$97.92 \\
\reactPro & - & - & \$18.61 & \$97.92 & \$116.53 \\
\rowcolor{gray!10}
\reactFL & - & - & \$0.96 & \$97.92 & \$98.88 \\
\bottomrule
\end{tabular}
\end{threeparttable}
}

\end{table}

\newpage

\PP{Fuzzing Cost vs. LLM Cost}
Since both large-scale fuzzing
and LLM usage
can incur significant costs,
we analyze
the cost breakdown
of \sys's configurations
and compare them
to baseline fuzzing costs
(\autoref{t:costs}).
To estimate fuzzing costs,
we reference AWS pricing~\cite{aws-ec2-pricing}
for comparable instance types
(c7a.16xlarge, 64-core AMD EPYC 4th gen),
which costs 3.28448\USD per hour, or 0.05132\USD per CPU hour.
Calculating for our three-CPU, 12-hour runs
(including baseline),
the fuzzing cost per harness
is 1.84752\USD,
resulting in a total fuzzing cost
of 97.92\USD
across 53 harnesses.
For the large-scale baseline
with 50 CPUs and 24 hours,
the fuzzing cost per harness
is 61.584\USD,
leading to a total fuzzing cost
of 3,263.95\USD
for all 53 harnesses.
We list these fuzzing costs
in \autoref{t:costs}
for comparison
with \sys's LLM costs.

\autoref{t:costs} shows that,
for flagship models,
\sys's LLM costs
significantly exceed
the fuzzing costs.
For example,
\sysGPT
incurs 1,369.24\USD
for sink exploration
and 608.28\USD for sink exploitation,
compared to only 97.92\USD
for fuzzing.
However,
even with high LLM costs,
\sys remains more cost-effective than large-scale fuzzing
while finding over three times more vulnerabilities,
indicating that investing in LLM usage
yields significantly better returns
than merely scaling fuzzing resources.

\finding{\sys costs less than large-scale fuzzing (\$2,429--\$3,051 vs.\ \$3,264) while exploiting 4--5\x more vulnerabilities (37--41 vs.\ 8).}


\PN{Model Cost--Performance Trade-offs.}
\autoref{t:costs} also reveals three distinct cost tiers.
Flagship models (GPT-5, Gemini-2.5-Pro, Sonnet-4.5)
exploit 37--41 vulnerabilities
at \$2,429--\$3,051 total,
yielding \$66--\$74 per exploited vulnerability.
Lightweight models (GPT-5-nano, Gemini-2.5-Flash-Lite)
solve 27/26 vulnerabilities
at \$182/\$208,
approximately
13\x/15\x cheaper than their corresponding flagships,
with
extra failures stemming from weaker reasoning and tool-use capabilities.
Open-source models offer a compelling middle ground:
GLM-5 exploits 35 vulnerabilities at only \$392 total
(\$11.21 per bug),
achieving within 2--6 vulnerabilities of flagship performance
at roughly 8\x lower cost.
GPT-OSS-120b is the cheapest option at \$5.11 per bug,
though it finds fewer vulnerabilities~(29).
These results confirm that \sys's design
naturally benefits from stronger LLMs
but remains effective with budget-friendly alternatives,
allowing practitioners to select models
based on their cost--performance requirements.
Costs can be further reduced
through per-sink incremental analysis in CI/PR workflows,
re-analyzing only sinks affected by code changes
rather than the entire project.

\finding{\sys's effectiveness varies with the underlying LLM: flagship models exploit 37--41 vulnerabilities versus 26--27 for lightweight models, at 12--17\x higher cost.}

\finding{Open-weight models achieve near-flagship effectiveness at a fraction of the cost: GLM-5 exploits 35 vulnerabilities at \$392 (8\x cheaper than flagships), offering a viable option for cost-sensitive deployments or privacy-constrained environments that require locally hosted models.}

\PP{Static Analysis Comparison}%
\autoref{t:sa-sota} compares \sys's static analysis pipeline
against CodeQL and SpotBugs
on all 8,262 candidate sinks
(54 exploitable, 8,208 non-exploitable).
Although our pipeline's precision is low (7--15.7\%),
this is by design:
missing an exploitable sink means missing a vulnerability entirely,
so we deliberately optimize for recall (92.6--96.3\%).
Established static analysis tools
exhibit even lower precision.
CodeQL achieves only 6.1\% precision
at 25.9\% recall,
missing 40 of the 54 exploitable sinks.
SpotBugs reaches 37\% recall
at 7.6\% precision,
still missing 34 of 54 exploitable sinks.

\PN{Vulnerabilities Uniquely Found by \sys.}
We found that \sysGemini exploits 14 out of the 21 vulnerabilities
that Jazzer reaches but does not exploit, due to \sys's
ability to handle complex input formats.
Ten of these vulnerabilities
require strict input formats such as archives (7),
XML (2), or domain-specific formats (1),
making them challenging to generate with a generic fuzzer like Jazzer.
\sys also exploits a stateful vulnerability
that Jazzer cannot handle
requiring multiple interactions
to reach the vulnerable state.
In two cases,
\sys overcomes the lack of sanitizer feedback
that hinders Jazzer from creating inputs
that contain the correct sentinels to trigger the sanitizer.
Finally,
\sys successfully exploits
a hash value computation vulnerability,
which is a notoriously difficult constraint
for fuzzers to solve.
Overall,
this demonstrates \sys's strength
in handling complex input formats
and scenarios, enabling it to discover
vulnerabilities
that Jazzer does not detect.

\begin{table}[t]
\centering
\caption{\sys sink detection vs.\ static analysis tools.}
\label{t:sa-sota}
\begin{threeparttable}
\scriptsize
\setlength{\tabcolsep}{3pt}
\begin{tabular}{l r r r r >{\raggedleft\arraybackslash}p{3.5em} r r r r}
\toprule
\textbf{Tool} & \textbf{TP} & \textbf{FP} & \textbf{FN} & \textbf{TN} & \textbf{Cost} & \textbf{Prec} & \textbf{Recall} & \textbf{F1} & \textbf{Spec} \\
\midrule
\GsysGPT & 52 & 331 & 2 & 7877 & \$353.39 & 13.6\% & 96.3\% & 23.8\% & 96.0\% \\
\rowcolor{gray!10}
\GsysGemini & 50 & 308 & 4 & 7900 & \$123.74 & 14.0\% & 92.6\% & 24.3\% & 96.2\% \\
\GsysSonnet & 51 & 304 & 3 & 7904 & \$751.93 & 14.4\% & 94.4\% & 24.9\% & 96.3\% \\
\rowcolor{gray!10}
\GsysNano & 51 & 413 & 3 & 7795 & \$4.60 & 11.0\% & 94.4\% & 19.7\% & 95.0\% \\
\GsysFlashLite & 51 & 273 & 3 & 7935 & \$40.99 & 15.7\% & 94.4\% & 27.0\% & 96.7\% \\
\rowcolor{gray!10}
\GsysGLM & 51 & 309 & 3 & 7899 & \$19.32 & 14.2\% & 94.4\% & 24.6\% & 96.2\% \\
\GsysGPTOSS & 52 & 692 & 2 & 7516 & \$1.74 & 7.0\% & 96.3\% & 13.0\% & 91.6\% \\
\midrule
\rowcolor{gray!10}
CodeQL & 14 & 215 & 40 & 7993 & \$1.13\tnote{$\dagger$} & 6.1\% & 25.9\% & 9.9\% & 97.4\% \\
SpotBugs & 20 & 244 & 34 & 7964 & \$1.13\tnote{$\dagger$} & 7.6\% & 37.0\% & 12.6\% & 97.0\% \\
\bottomrule
\end{tabular}
\begin{tablenotes}
\setlength{\itemindent}{-10pt}
\item \textbf{TP/FP/FN/TN}: True/False Positives/Negatives out of 54 exploitable and 8,208 non-exploitable sinks
\item \textbf{Prec}: Precision, \textbf{Spec}: Specificity
{\setlength{\itemindent}{-5pt}\item[$\dagger$] Estimated using AWS pricing (as for \autoref{t:costs}), 1 cpu-hour per project}
\end{tablenotes}
\end{threeparttable}

\end{table}

We also find that \sys is able to reach and exploit
21 out of the 25 vulnerabilities
that even \baselineLS cannot reach.
Similar to the previous case,
the majority (17) of these vulnerabilities
require complex input formats
such as archives (5),
XML (4), or domain-specific formats (8) like Java class files, RTF, and SVG.
On top,
two stateful vulnerabilities
are missed by Jazzer
and another two remain unreachable
due to lack of instrumentation.
\sys overcomes these challenges
by leveraging LLMs to generate
semantically valid inputs
even in cases where complex formats
or stateful interactions are required.



\PP{Vulnerabilities Missed in All Configurations}
We find that the remaining 11 vulnerabilities
that \sys misses across all configurations
primarily stem from the complexity of their input formats.
Specifically,
four vulnerabilities
require non-standard modifications
to common file formats,
which are challenging for the LLMs to generate correctly.
Three vulnerabilities
necessitate stateful interactions while involving complex input formats at the same time,
posing a dual challenge.
Two vulnerabilities
involve combinations of different input formats (such as image data embedded in an HTML file),
which complicates input generation.
Finally,
two vulnerabilities
require more complex Java object serialization (such as multiple serialized objects)
that are difficult for current LLMs to handle.
These challenges highlight
the limitations of current LLMs
in generating highly complex or stateful inputs.
However,
we believe
that these are not fundamental limitations of \sys's approach.
With advancements in LLM capabilities
or the integration of specialized tools
for specific CWE types,
\sys could potentially overcome these challenges
in future work,
enabling it to discover
even more complex vulnerabilities.

\finding{\sys exploits 35 of 46 vulnerabilities that Jazzer misses by leveraging LLM semantic reasoning to satisfy complex constraints that mutation-based fuzzing cannot address.}



\subsection{RQ2: Component Effectiveness}

\PP{Ablation Setup}
We compare \sys with individual components disabled
against the baselines and full settings, as described
in the Evaluation Setup.
We add the following ablation configurations
to the comparison, which all use GPT-5 as the LLM:

\begin{itemize}
    \item \textbf{\sys-RO}: Disabled sink exploitation agent (\textbf{R}eachability \textbf{O}nly)
    \item \textbf{\sys-XO}: Disabled sink exploration agent (e\textbf{X}ploitation \textbf{O}nly)
    \item \textbf{\sys-AF}: Union of sink exploration and sink exploitation \textbf{A}gent \textbf{F}indings directly; sink exploitation output is taken from the \sysGPT configuration
\end{itemize}

We evaluate these configurations
on the same metrics for reached
and exploited vulnerabilities
and show results
in the same diagram
(\autoref{f:tool-coords}),
per-vuln overview
(\autoref{f:cpv-matrix}),
and CWE-based breakdowns
(\autoref{f:spider-per-config}, (iv)-(vi)).

\PP{Contribution of Sink Exploration and Exploitation}
Comparing \sys-RO and \sys-XO with the full \sys
configuration
shows how the sink exploration and exploitation contribute
a significant number of reached
and exploited vulnerabilities (\autoref{f:cpv-matrix}).
First, disabling the sink exploration agent
dramatically reduces
the number of reached vulnerabilities
compared to the full setting
(29 vs. 42).
Visually, this is reflected
in a significantly smaller overall area
in the spider graph of \sys-XO (Graph (v) in \autoref{f:spider-per-config})
demonstrating the importance of the sink exploration agent.
Second, disabling the sink exploitation agent
(\sys-RO, Graph (iv) in \autoref{f:spider-per-config})
substantially reduces
the number of exploited vulnerabilities
compared to the full setting
(18 vs. 37),
illustrated by a significantly smaller dark gray area
compared to the full \sys configuration of \sysGPT (Graph (i)).
Overall, this demonstrates that
both the sink exploration and sink exploitation agents
are critical
for reaching vulnerabilities and exploiting reached ones, respectively.

\PP{Mutual Benefits of Agent-Fuzzer Cooperation}
Comparing the union of the agent-only findings and
the fuzzing results (\sys-AF and \baseline)
with the full \sys configurations
shows that \sys performs significantly better
not because of the individual components alone,
but due to their interplay.
\autoref{f:cpv-matrix}
reveals that seven vulnerabilities
are uniquely discovered
by the combined approach
and not detected
by either the fuzzing baseline or the agent-only setting alone.
This is because the LLM helps as an unblocker
for certain paths
where the fuzzer is stuck,
while the fuzzer
can continue to rapidly mutate inputs
to finally reach or exploit a vulnerability
that the agent generates a semantically close (but not sufficient) input for.
These synergy-only vulnerabilities
validate \sys's cooperative design:
agents contribute semantically grounded inputs
that the fuzzer refines through mutation,
while the fuzzer provides beep seeds
that ground the agents' exploitation reasoning.
The combination
of LLM agents and fuzzing
significantly outperforms
the union of their individual findings.

\finding{Agent-fuzzer cooperation discovers 7 vulnerabilities that neither component finds alone, confirming that \sys's synergistic design outperforms the sum of its parts.}

\PP{Case study: CWE-022 (Path Traversal)}
This vulnerability class exemplifies how Jazzer and the LLM agents work together.
Analysis of the spider graphs from our ablation study (\autoref{f:spider-per-config}) yields three key findings.

First, the \sys-RO results
highlight the challenges
of exploiting certain vulnerabilities
even after reaching them.
Compared to the full \sys configurations,
\sys-RO exploits significantly fewer path traversals
(2 vs. 8).
We investigate these cases in detail
and find that many require structured inputs
(e.g., valid file formats)
to trigger Jazzer's sanitizers.
For example, the BCEL vulnerability
requires a valid Java class file as input,
while the Widoco and ZTZIP vulnerabilities
require valid ZIP files.
Similarly,
the Jenkins path traversal vulnerability
necessitates constructing a specific program state,
which depends on structured inputs as well.
When Jazzer mutates these inputs,
it often breaks the required structure,
preventing it from reaching the sink with the mutated inputs.
This demonstrates the last-mile challenge of exploitation
and underscores the importance of the exploitation agent.

Second, the \sys-XO results in \autoref{f:spider-per-config}
emphasize the critical role
of the sink exploration agent
in reaching certain vulnerabilities.
Compared to the full \sys configurations,
\sys-XO reaches significantly fewer path traversals (5 vs. 10), 
and consequently only exploits those five.
Investigating these cases,
we find that \sys-XO has difficulty reaching the same vulnerabilities
that \sys-RO fails to exploit.
However, this time, Jazzer is not even able to reach the sinks
with its mutations,
since it fails to craft the required structured inputs
(e.g., valid file formats or program states).
This further demonstrates
the limitations of Jazzer's mutation strategy
in reaching complex vulnerabilities,
highlighting the importance
of the exploration agent in reaching these challenging sinks.


\subsection{RQ3: Agent Analysis}

We evaluate LLM agent performance
across seven models for each module:
classification accuracy for sink filtering,
iteration efficiency for sink exploration,
and exploit synthesis success for sink exploitation.

\PP{Sink Filtering}%
The sink filtering results in \autoref{t:agent-filtering}
confirm our design decision
to prioritize recall over precision
while reducing the need for downstream analysis resources.
With any model, the agent achieves over 96\% recall
(50--52 of 52 exploitable sinks retained),
ensuring that nearly all true vulnerabilities
enter the dynamic analysis pipeline.
We attribute the high recall to
the bounded task scope
of harness reachability and sink specifics
reducing classification complexity
for the LLMs.
The only false negatives stem from
incomplete data-flow collection
(five models on one sink)
or misunderstood variable controllability
(Gemini-2.5-Pro on another).
At the same time,
most configurations filter out 55--70\%
of non-exploitable sinks,
substantially reducing the downstream search space;
the exception is GPT-OSS-120B,
which struggles to use the filtering tools effectively.

\finding{LLM-based sink filtering retains $>$96\% of exploit\-able sinks while removing 23--66\% of all sinks, substantially reducing the downstream search space at minimal miss rate.}

\begin{table}[t]
\centering
\caption{Sink filtering agent effectiveness across models.}
\label{t:agent-filtering}
\begin{threeparttable}
\scriptsize
\setlength{\tabcolsep}{3pt}
\begin{tabular}{l r r r r >{\raggedleft\arraybackslash}p{3.5em} r r r r}
\toprule
\textbf{Model} & \textbf{TP} & \textbf{FP} & \textbf{FN} & \textbf{TN} & \textbf{Cost} & \textbf{Prec} & \textbf{Recall} & \textbf{F1} & \textbf{Spec} \\
\midrule
\GsysGPT & 52 & 331 & 0 & 578 & \$353.39 & 13.6\% & 100.0\% & 23.9\% & 63.6\% \\
\rowcolor{gray!10}
\GsysGemini & 50 & 308 & 2 & 601 & \$123.74 & 14.0\% & 96.2\% & 24.4\% & 66.1\% \\
\GsysSonnet & 51 & 304 & 1 & 605 & \$751.93 & 14.4\% & 98.1\% & 25.1\% & 66.6\% \\
\rowcolor{gray!10}
\GsysNano & 51 & 413 & 1 & 496 & \$4.60 & 11.0\% & 98.1\% & 19.8\% & 54.6\% \\
\GsysFlashLite & 51 & 273 & 1 & 636 & \$40.99 & 15.7\% & 98.1\% & 27.1\% & 70.0\% \\
\rowcolor{gray!10}
\GsysGPTOSS & 52 & 692 & 0 & 217 & \$1.74 & 7.0\% & 100.0\% & 13.1\% & 23.9\% \\
\GsysGLM & 51 & 309 & 1 & 600 & \$19.32 & 14.2\% & 98.1\% & 24.8\% & 66.0\% \\
\bottomrule
\end{tabular}
\begin{tablenotes}
\setlength{\itemindent}{-10pt}
\item \textbf{TP/FP/FN/TN}: True/False Positives/Negatives out of 52 exploitable and 909 non-exploitable sinks
\item \textbf{Prec}: Precision, \textbf{Spec}: Specificity
\end{tablenotes}
\end{threeparttable}

\end{table}

\PP{Sink Exploration}%
\sys's iterative exploration design
helps weaker models narrow the gap
with flagship models
through additional reasoning rounds
(\autoref{t:agent-exploration}).
Flagship models solve 47--53 of the 88 call-paths,
with 42--48 already solved on the first generated input.
Open-weight and lightweight models
start with fewer first-input successes (28--35),
but iterative refinement helps narrow this gap:
GLM-5 solves only 32 call-paths on the first input
yet reaches 44 after iterations,
gaining 12 additional call-paths
and approaching flagship performance.

To understand why certain call-paths remain unsolved,
we manually classified \sysGemini's attempts
(\autoref{s:agent-classification}).
Solved call-paths are resolved
through localized code context and common-sense knowledge,
such as well-known input formats.
Unsolved ones are dominated
by static analysis barriers,
particularly indirect calls,
which prevent the agent
from collecting sufficient path context.

\begin{table}[t]
\centering
\caption{Exploration agent: inputs generated per call-path.}
\label{t:agent-exploration}
\begin{threeparttable}
\scriptsize
\setlength{\tabcolsep}{3pt}
\begin{tabular}{l >{\raggedleft\arraybackslash}p{3em} >{\raggedleft\arraybackslash}p{3em} >{\raggedleft\arraybackslash}p{3em} >{\raggedleft\arraybackslash}p{3em} >{\raggedleft\arraybackslash}p{3em} >{\raggedleft\arraybackslash}p{3em}}
\toprule
 & \multicolumn{3}{c}{\textbf{Reached}} & \multicolumn{3}{c}{\textbf{Not Reached}} \\
\cmidrule(lr){2-4} \cmidrule(lr){5-7}
\textbf{Model} & \textbf{1} & \textbf{2} & \textbf{3+} & \textbf{1} & \textbf{2} & \textbf{3+} \\
\midrule
\GsysGPT & 42 & 4 & 1 & 0 & 23 & 18 \\
\rowcolor{gray!10}
\GsysGemini & 48 & 4 & 1 & 0 & 2 & 33 \\
\GsysSonnet & 44 & 3 & 0 & 0 & 1 & 40 \\
\rowcolor{gray!10}
\GsysNano & 32 & 2 & 0 & 0 & 17 & 37 \\
\GsysFlashLite & 28 & 7 & 0 & 0 & 10 & 43 \\
\rowcolor{gray!10}
\GsysGPTOSS & 35 & 6 & 0 & 0 & 5 & 42 \\
\GsysGLM & 32 & 11 & 1 & 0 & 32 & 12 \\
\bottomrule
\end{tabular}
\begin{tablenotes}
\setlength{\itemindent}{-10pt}
\item Number of inputs generated before the agent solved or failed each call-path (88 total)
\item \textbf{Reached/Not Reached}: whether a generated input reached the target sink
\end{tablenotes}
\end{threeparttable}

\end{table}

\PP{Sink Exploitation}%
Exploitation success depends on both
iterative refinement and model capability
(\autoref{t:agent-exploitation}).
At most 38\% of attempts succeed
with the first generated input,
yet flagship models achieve 61--70\% overall success.
Beside multi-round refinement, the No-Coverage Fuzz step (\autoref{s:design:exploitation})
is a key driver of this improvement,
converting 17--43\% of the cases
where the agent itself failed.
Beyond iteration depth,
model capability also matters:
flagship models significantly outperform
lightweight counterparts (29--47\% success).

To understand the root causes
of exploitation successes and failures,
we manually classified \sysGemini's and \sysFlashLite's attempts
(\autoref{s:agent-classification}).
Successful exploits primarily overcome
input formatting issues,
such as grammar violations
and script coding errors.
Failures, in contrast, stem from
complex serialized or project-specific input formats
and unsatisfied sanitizer conditions,
pointing to the fundamental challenge
of generating semantically valid inputs
for highly structured formats.

\finding{Iterative input generation is essential: at most 38\% exploitation attempts succeed with the first input, yet achieve 61--70\% overall success through multi-round refinement and no-coverage fuzzing (numbers for flagship models).}

\begin{table}[t]
\centering
\caption{Exploitation agent: inputs generated per exploitation attempt.}
\label{t:agent-exploitation}
\begin{threeparttable}
\scriptsize
\setlength{\tabcolsep}{1.5pt}
\begin{tabular}{l S[table-format=1.1,mode=text] S[table-format=2.1,mode=text] S[table-format=1.1,mode=text] S[table-format=2.1,mode=text] >{\raggedleft\arraybackslash}p{3em} >{\raggedleft\arraybackslash}p{3em} >{\raggedleft\arraybackslash}p{3em} >{\raggedleft\arraybackslash}p{3em} >{\raggedleft\arraybackslash}p{3.5em} r}
\toprule
 & \multicolumn{2}{c}{\textbf{Beepseed}} & \multicolumn{2}{c}{\textbf{Path}} & \multicolumn{5}{c}{\textbf{Inputs Generated (Vulns)}} &  \\
\cmidrule(lr){2-3} \cmidrule(lr){4-5} \cmidrule(lr){6-10}
\textbf{Model} & {\textbf{Avg}} & {\textbf{P90}} & {\textbf{Avg}} & {\textbf{P90}} & \textbf{1} & \textbf{2--4} & \textbf{5+} & \textbf{NoCov} & \textbf{Fail} & \textbf{Succ} \\
\midrule
\GsysGPT & 4.6 & 10.6 & 3.3 & 9.0 & 29 (20) & 14 (6) & 7 (2) & 36 (4) & 43 (20) & 61\% \\
\rowcolor{gray!10}
\GsysGemini & 3.7 & 6.8 & 2.1 & 3.0 & 16 (12) & 13 (4) & 6 (3) & 22 (9) & 44 (12) & 70\% \\
\GsysSonnet & 4.1 & 12.4 & 2.9 & 10.6 & 11 (8) & 19 (7) & 20 (11) & 35 (5) & 41 (20) & 61\% \\
\rowcolor{gray!10}
\GsysNano & 5.9 & 23.5 & 3.6 & 18.4 & 7 (6) & 7 (6) & 6 (5) & 30 (7) & 71 (27) & 47\% \\
\GsysFlashLite & 4.7 & 10.8 & 2.4 & 6.6 & 4 (4) & 4 (2) & 1 (1) & 30 (8) & 79 (36) & 29\% \\
\rowcolor{gray!10}
\GsysGPTOSS & 4.8 & 6.7 & 2.6 & 3.7 & 14 (13) & 3 (3) & 1 (1) & 17 (5) & 127 (18) & 55\% \\
\GsysGLM & 5.6 & 8.0 & 2.7 & 3.0 & 7 (7) & 2 (2) & 0 (0) & 7 (5) & 195 (25) & 36\% \\
\bottomrule
\end{tabular}
\begin{tablenotes}
\setlength{\itemindent}{-10pt}
\item \textbf{Beepseed/Path}: number of beepseeds/paths per sinkpoint, \textbf{Avg}: average, \textbf{P90}: 90th percentile
\item \textbf{NoCov}: No-coverage fuzzing solves (Alg.\,2, line\,11), \textbf{Succ}: overall success rate
\end{tablenotes}
\end{threeparttable}

\end{table}

\subsection{RQ4: Industry Validation and Adoption}

\PP{AIxCC Performance}%
The DARPA AI Cyber Challenge (AIxCC)~\cite{aixcc}
provided a rigorous third-party setting
to assess \sys's real-world effectiveness,
with an independent evaluation and unseen challenges.
During the AIxCC final competition,
autonomous systems were evaluated
on real-world open-source C and Java software
with DARPA-injected vulnerabilities.
DARPA and its contractors designed the challenges,
operated the competition infrastructure, and scored the results,
investing tens of thousands of dollars
in computational resources;
no team had prior access to~the~challenges.

On the Java challenges,
an early version of \sys's techniques
formed a major part of the Java vulnerability discovery module (Atlantis-Java)
in Team Atlanta's first-place CRS~\cite{zhang2026sokaixcc},
which contributed the majority of Java vulnerability discoveries
in the competition.
This early version shared \sys's core sink-centric design,
including sink detection
and the collaborative exploration and exploitation agents;
the version presented in this paper
adds a more systematic sink detection pipeline
and refines the exploitation agent.
Specifically, \sys discovered
seven DARPA-injected, sink-based vulnerabilities
across three CWE categories
(four CWE-022, two CWE-078, and one CWE-074)
and three zero-day vulnerabilities
in Hertzbeat, Healthcare-Data-Harmonization, and PDFBox.
These results independently confirm
that our sink-centric approach
is effective on real-world software
beyond our benchmark.

\PP{OpenSSF Collaboration}%
Following Team Atlanta's first-place finish,
the OpenSSF reached out to collaborate
on applying the team's CRS
to secure open-source software at scale.
This led to OSS-CRS~\cite{oss-crs},
a sandbox project\footnote{\href{https://github.com/ossf/tac/blob/main/process/project-lifecycle.md\%23sandbox}{https://github.com/ossf/tac/blob/main/process/project-lifecycle.md\#sandbox}} in the OpenSSF,
into which Team Atlanta's CRS is integrated\footnote{\url{https://github.com/ossf/oss-crs/blob/689a2451dacc13a12deed6f80fdab18b9ddf56a2/registry/atlantis-java-main.yaml}}
for analyzing open-source projects.
As a registered, standalone Java vulnerability detection CRS in OSS-CRS,
\sys has already found a zero-day path traversal vulnerability
in a common Java database implementation, which the maintainer
has confirmed and is patching (anonmized for responsible disclosure).
This early deployment and the zeroday finding demonstrate that \sys
transfers beyond our benchmark
to real-world open-source software.

\section{Discussion}
\label{s:discussion}

\PP{Fuzzing Harness as Program Entrypoint}
Like other fuzzing-based approaches,
\sys requires fuzzing harnesses to serve as program entry points
for both dynamic and static analysis components,
aligning with standard fuzzing practice.
This means a functional harness
must reach the target sink's enclosing method,
just like any dynamic proof-of-vulnerability generation approach
including Jazzer;
if a harness covers limited code,
\sys will miss vulnerabilities
in the unreachable portions.
However, \sys's full compatibility with the OSS-Fuzz project format
mitigates this limitation:
it can leverage all existing harnesses
developed by the open-source community
and, since \sys is orthogonal to automatic harness generation techniques
such as OSS-Fuzz-Gen~\cite{oss-fuzz-gen},
it can directly benefit from them
to broaden coverage
to projects that currently lack manual harnesses.
Moreover, the harness-based approach brings a key benefit:
\sys generates truly exploitable,
verifiable proof-of-concept inputs
rather than theoretical vulnerability reports.

\PP{Extending to Other CWEs and Languages}
\sys's framework is designed with cross-CWE generality in mind,
minimizing the effort required to support new sink-based vulnerability types.
The sink detection component achieves this
through a naturally extensible design
that maximizes reuse of CodeQL's sink database,
query scripts, and infrastructure.
When adding support for a new CWE type,
the core framework remains unchanged;
only three specific artifacts require updates:
the CWE-to-sink mapping query script in CodeQL,
and the CWE-specific descriptive text
in both the exploration and exploitation agents.

Beyond CWE extensibility,
our approach can theoretically extend to other programming languages
such as \cc{C}/\cc{C++}, \cc{Python}, and \cc{Go}.
However, such extensions would require language-specific adaptations.
The primary requirement is modifying
the coverage-guided fuzzer for the target language
to support beep seed collection.
Additionally,
each language can present unique static/dynamic analysis challenges in \cc{Joern} or \cc{CodeQL} settings or debugger integration.
While these challenges are non-trivial,
they present language-specific engineering efforts
rather than fundamental limitations of \sys's design.

\PP{Incorporating Advanced CWE-Specific Techniques}
The current \sys implementation deliberately avoids
CWE-specific optimization techniques,
instead focusing on general-purpose agent capabilities
that work across multiple vulnerability types.
This design choice demonstrates
the effectiveness of the core framework
without relying on specialized tools.
However, the agent-based architecture
naturally accommodates per-CWE enhancements
when additional effectiveness is desired.

Specifically,
agents can be augmented with additional tools and specialized prompts
for particular CWE types they support.
For example,
the exploitation agent could integrate
gadget chain search~\cite{gadget-search}~\cite{oddfuzz}~\cite{ysoserial}
to enhance unsafe deserialization exploit generation.
For ReDoS vulnerabilities,
the agent could incorporate
rule-based tools~\cite{recheck} that assess
whether a potentially vulnerable regex pattern
is actually exploitable
and generate concrete exploit strings.
These CWE-specific enhancements
would complement \sys's framework,
providing targeted performance improvements.

\PP{Benchmark Contamination Risk}
A potential concern when evaluating LLM-based systems
is data contamination:
models may have seen vulnerability details during training,
inflating exploitation success.
By construction, our benchmark is largely free of contamination risk
(\autoref{t:contamination} in the appendix).
32 of 54 challenges carry no contamination risk:
15 are self-synthetic,
and 17 are from AIxCC pre-final challenges,
which remain unpublished
and were not available for model training before August 2025
according to the competition organizers,
post-dating the cutoffs of all evaluated models
with a disclosed cutoff date (\autoref{t:models}).
The remaining 22 are CVE-based or exemplar-derived,
where only metadata (descriptions, patches)
or different-context PoVs are public;
only 1 of 54 has a potentially similar public PoV.
To further mitigate contamination risk going forward,
we will distribute the benchmark
only under gated access to researchers and practitioners,
minimizing the risk of vulnerability details
being included in future LLM training~data.

Beyond composition,
our experiments show no empirical signs
that contamination drives \sys's results.
Agent logs show no evidence of memorization:
no CVE-specific identifiers
or unique string patterns from public exploits
appear in any agent trace.
In addition,
the ReAct baselines use the same LLMs
but perform far below \sys
(12--16 vs.\ 26--41 exploited),
showing that gains stem from
\sys's framework design
rather than model memorization.

\section{Related Work}
\label{s:relwk}


\PP{Dynamic Vulnerability Discovery in Memory-Safe Language}
While memory-safe languages like Java, Python, PHP, and Golang
eliminate traditional memory corruption vulnerabilities,
they still suffer from logical vulnerabilities.
Coverage-guided fuzzers such as
Jazzer~\cite{jazzer}, JQF~\cite{jqf},
Atheris~\cite{atheris}, and go-fuzz~\cite{go-fuzz},
inherit their design from C/C++ memory corruption testing,
treating all code paths equally
without prioritizing security-sensitive sinks.

Many approaches focus on specific vulnerability types,
including deserialization~\cite{rasheed2020hybrid,oddfuzz,gadget-search,cao2023overriding,haken2018gadgets},
algorithmic complexity and DoS~\cite{acquirer,hotfuzz},
injection vulnerabilities~\cite{witcher},
file upload~\cite{uchecker,fuse},
SSRF~\cite{ssrf-discovery},
prototype pollution~\cite{uopf,li2021objlupansys,silentspring},
and cross-thread vulnerabilities~\cite{jaex}.
Those are CWE-specific techniques which tackle challenges unique to each vulnerability type.
For instance, for deserialization vulnerabilities,
JDD~\cite{gadget-search} proposed a bottom-up gadget search approach
to address path explosion in Java object injection detection,
while OddFuzz~\cite{oddfuzz} and FUGIO~\cite{fugio} further improved
gadget chain mining and exploit generation~\cite{cao2023overriding,haken2018gadgets,rasheed2020hybrid}.
These techniques are complementary to \sys,
as their specialized techniques
can be integrated to enhance CWE-specific detection.

Beyond CWE-specific approaches,
several works have also explored general sink-aware fuzzing frameworks,
including Witcher~\cite{witcher}, webFuzz~\cite{webfuzz}, Atropos~\cite{atropos},
WDFuzz~\cite{wdfuzz}, Predator~\cite{predator}, and BackREST~\cite{backrest}.
However, these approaches rely primarily on
traditional program analysis techniques
such as taint tracking, coverage feedback, and directed scheduling.
While effective at structural understanding
like identifying data flow paths and reachability constraints,
they struggle to leverage semantic sink knowledge,
such as API usage semantics and exploitation conditions.
\sys uniquely combines structural program analysis with semantic reasoning
to systematically extract and utilize sink-specific knowledge.


\PP{LLM for Fuzzing}
Recent advancements in LLMs
have opened new opportunities for fuzzing research,
primarily focusing on two areas:
input generation and driver generation.
For input generation,
approaches like TitanFuzz~\cite{deng2023titanfuzz},
Fuzz4All~\cite{xia2024fuzz4all}, and CODAMOSA~\cite{lemieux2023codamosa}
directly generate test cases for specific domains,
while SeedMind~\cite{shi2024seedmind} synthesizes seed generators iteratively.
Other works focus on generating fuzzing components:
ELFuzz~\cite{chen2025elfuzz} synthesizes entire fuzzers via LLM-driven evolution,
ChatFuzz~\cite{hu2023chatfuzz} and ChatAFL~\cite{meng2024chatafl}
leverage LLMs for structure-aware mutation and protocol fuzzing,
and WhiteFox~\cite{yang2024whitefox} and KernelGPT~\cite{yang2025kernelgpt}
apply LLMs to compiler and kernel fuzzing.
For driver generation,
works like OSS-Fuzz-Gen~\cite{liu2024ossfuzzgen},
Zhang's~\cite{zhang2024effective},
and PromptFuzz~\cite{lyu2024promptfuzz}
leverage LLM code-writing capabilities
to scale fuzz driver synthesis for library APIs.
These input generation approaches are orthogonal to \sys,
as they focus on general input structure and format,
while \sys focuses on vulnerability-specific knowledge contextualization
for sink-based guidance.
Driver generation approaches are complementary to \sys
and can be combined for more comprehensive automated vulnerability discovery.

\section{Conclusion}
\label{s:conclusion}

We presented \sys,
a sink-centric fuzzing framework
that systematically leverages sink knowledge
through collaborative integration of LLMs, program analysis, and fuzzing.
By identifying high-potential sinks,
guiding exploration toward vulnerable code,
and developing targeted exploits,
\sys exploits 41 vulnerabilities compared to Jazzer's 8
(a 4\x improvement)
on our benchmark of \numcpvs vulnerabilities across \numcwes CWE types.
Beyond benchmark evaluation,
\sys contributed to Team Atlanta's winning CRS
in the DARPA AIxCC
and is integrated into OSS-CRS,
a sandbox project in the OpenSSF,
validating its practical impact
for protecting real-world open-source software.

\section{Acknowledgment}
\label{s:ack}

We thank the anonymous reviewers
and our shepherd
for their helpful feedback.
We also thank
Ammar Askar,
Dae R. Jeong,
Ding Zhang,
Dohyeok Kim,
Hyungseok Han,
James Yao,
Jiho Kim,
Junsik Kim,
Kyungjoon Ko,
Leyan Pan,
Sangwoo Ji,
Saumya Agarwal,
Soyeon Park,
Wonyoung Kim,
Yeongjin Jang,
and Youngjoon Kim
for their contributions to the benchmark construction.

We thank DARPA and the AIxCC organizers
for designing and running the competition,
and all teams
for the competitive environment
that drove innovation.

This work was supported by the Institute of Information \&
Communications Technology Planning \& Evaluation (IITP) grant funded by
the Korea government (MSIT) (No.RS-2024-00440780, Development of
Automated SBOM and VEX Verification Technologies for Securing Software
Supply Chains),
the Advanced Research Projects Agency for Health (ARPA-H) under Other Transaction Agreement No. 140D042590046,
FriendliAI through API credits,
and the AIxCC prize money received by Team Atlanta.

\section{Artifact Availability}
\label{s:artifact}

Our system implementation, datasets, and evaluation data
are available at \url{https://doi.org/10.5281/zenodo.17607043}.
The source code of \sys is also part of
\url{https://github.com/Team-Atlanta/atlantis-java},
which integrates with OSS-CRS~\cite{oss-crs}
via \url{https://github.com/ossf/oss-crs/blob/689a2451dacc13a12deed6f80fdab18b9ddf56a2/registry/atlantis-java-main.yaml}.

\section{LLM Usage Considerations}
\label{s:llm-usage}

We used LLMs in two contexts.
First, LLMs were used for editorial purposes in this manuscript,
including refinement of main text, tables, and abstract,
but excluding references.
All outputs were inspected by the authors
to ensure accuracy and originality.

Second, our approach incorporates LLMs
as a core component of our fuzzing methodology.
We provide detailed discussions
of the necessity, limitations, effectiveness,
and technical details of LLM integration
in \autoref{s:design}, \autoref{s:eval}, and \autoref{s:discussion}.

\section{Ethics Considerations}
\label{s:ethics}

All discovered 0-day vulnerabilities from the third-party assessment
were responsibly disclosed
to the respective maintainers,
providing detailed reports, proof-of-concept inputs,
reproduction steps, and candidate patches.
We allowed sufficient time for remediation
before any public disclosure.

\bibliographystyle{ieeetr}
\bibliography{p,sslab,conf}

\appendices

\section{Benchmark Contamination Analysis}
\label{s:contamination}

\autoref{t:contamination} summarizes
the composition and contamination risk
of our benchmark.

\begin{table}[h]
\centering
\caption{Benchmark composition and contamination risk.}
\label{t:contamination}
\scriptsize
\setlength{\tabcolsep}{3pt}
\begin{tabular}{l r p{14em}}
\toprule
\textbf{Source} & \textbf{\#} & \textbf{Contamination Risk} \\
\midrule
Self-synthetic & 15 & None \\
\rowcolor{gray!10}
AIxCC pre-final (unpublished) & 17 & None \\
AIxCC exemplar videos (2024 Sep) & 3 & Metadata (video-based, no public source/PoVs) \\
\rowcolor{gray!10}
CVE-based (metadata only) & 15 & Metadata (descriptions, patches) \\
CVE-based (different payload) & 3 & PoV exists under different context \\
\rowcolor{gray!10}
CVE-based (similar PoV) & 1 & Potentially similar to public PoV \\
\midrule
\textbf{Total} & \textbf{54} & \\
\bottomrule
\end{tabular}
\end{table}

\section{Agent Manual Classifications}
\label{s:agent-classification}

We manually classified
the exploration attempts of \sysGemini
across all 88 call-paths
(\autoref{t:exploration-classification})
and the exploitation attempts of \sysGemini and \sysFlashLite
across 279 invocations
(\autoref{t:exploitation-classification}).

\begin{table}[h]
\centering
\caption{Manual classification of \sysGemini exploration attempts.}
\label{t:exploration-classification}
\scriptsize
\setlength{\tabcolsep}{3pt}
\begin{tabular}{l l >{\columncolor{white}}l r}
\toprule
\textbf{Outcome} & \textbf{Super-Category} & \textbf{Category} & \textbf{\#} \\
\midrule
\multirow{4}{*}{Solved (53)}
 & \multirow{2}{*}{Code context (30)} & Direct code references & 24 \\
 &  & Local semantics & 6 \\
\cmidrule(lr){2-4}
 & \multirow{2}{*}{Common sense (23)} & Well-known input formats & 19 \\
 &  & Base class instantiations & 4 \\
\midrule
\multirow{6}{*}{Unsolved (35)}
 & \multirow{4}{*}{Static analysis barriers (28)} & Indirect calls & 20 \\
 &  & Variable references & 3 \\
 &  & Class resolution & 3 \\
 &  & Inter-procedural dependencies & 2 \\
\cmidrule(lr){2-4}
 & \multirow{2}{*}{Complex input formats (7)} & FuzzedDataProvider & 5 \\
 &  & XML formats & 2 \\
\bottomrule
\end{tabular}
\end{table}

\begin{table}[h]
\centering
\caption{Manual classification of exploitation attempts (\sysGemini and \sysFlashLite).}
\label{t:exploitation-classification}
\scriptsize
\setlength{\tabcolsep}{3pt}
\begin{tabular}{l l >{\columncolor{white}}l r}
\toprule
\textbf{Outcome} & \textbf{Super-Category} & \textbf{Category} & \textbf{\#} \\
\midrule
\multirow{6}{*}{Solved (97)}
 & \multirow{2}{*}{Output formats (59)} & Grammar violations & 37 \\
 &  & Script coding errors & 22 \\
\cmidrule(lr){2-4}
 & \multirow{2}{*}{Exploit logic (15)} & Wrong exploit vectors & 10 \\
 &  & Incomplete exploit chains & 5 \\
\cmidrule(lr){2-4}
 & \multirow{2}{*}{Sanitizer (23)} & Unsatisfied sanitizer conditions & 14 \\
 &  & Wrong path-traversal payloads & 9 \\
\midrule
\multirow{6}{*}{Failed (182)}
 & \multirow{4}{*}{Complex input formats (122)} & Serialized objects & 55 \\
 &  & Project-specific formats & 43 \\
 &  & Bytecode & 15 \\
 &  & Nested formats & 9 \\
\cmidrule(lr){2-4}
 & \multirow{4}{*}{Sanitizer conditions (60)} & Deserialization & 22 \\
 &  & XPath injection & 15 \\
 &  & SSRF & 10 \\
 &  & Others & 13 \\
\bottomrule
\end{tabular}
\end{table}

\section{Jazzer Coverage Over Time}
\label{s:coverage}

\begin{figure}[h]
\centering
\includegraphics[width=\linewidth]{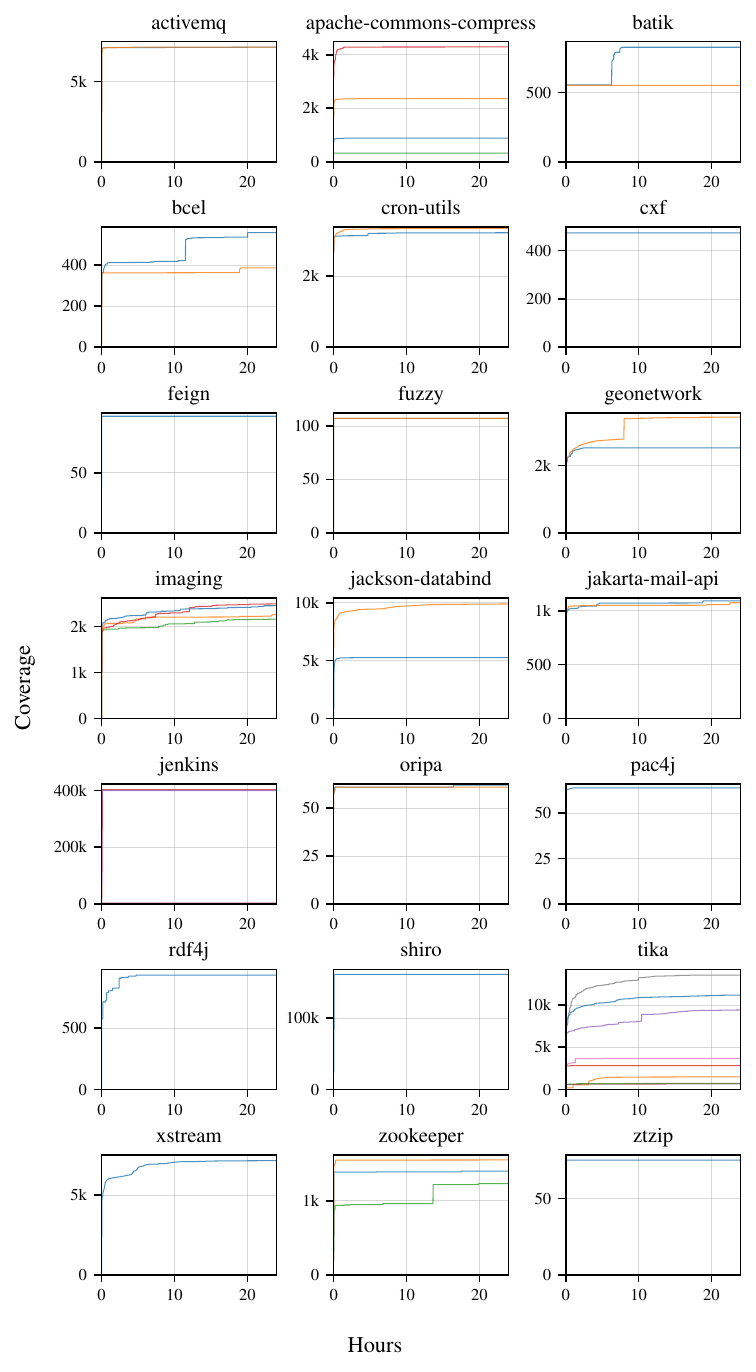}
\caption{Jazzer coverage over time per project.}
\label{f:coverage}
\end{figure}

\autoref{f:coverage} shows the coverage progression
of Jazzer across all benchmark projects over 24 hours.
Each line represents a single harness.

\clearpage

\section{Meta-Review}

The following meta-review was prepared by the program committee for the 2026
IEEE Symposium on Security and Privacy (S\&P) as part of the review process as
detailed in the call for papers.

\subsection{Summary}
The paper presents GONDAR, a sink-centric Java fuzzing framework that uses CodeQL to identify/filter sinks and two LLM agents to (i) generate sink-reaching inputs using call-path context and debugger feedback and (ii) refine them into sanitizer-triggering PoVs, optionally followed by short fuzzing. On a benchmark of 54 vulnerabilities, it reports substantially more exploited cases than Jazzer.

\subsection{Scientific Contributions}
\begin{itemize}
\item Creates a New Tool to Enable Future Science.
\item Provides a Valuable Step Forward in an Established Field.
\end{itemize}

\subsection{Reasons for Acceptance}
\begin{enumerate}
\item The paper provides a valuable step forward in an established field via a coherent end-to-end design targeting the reachability–exploitability gap in sink-based vulnerability discovery.
\item The paper creates a new tool to enable future science which provides substantial improvement over the chosen baseline, supported by ablations analysis.
\item The new tool offers potential community value if the framework and benchmark are released.
\end{enumerate}

\subsection{Noteworthy Concerns} 
\begin{enumerate} 
\item Internal precision is very low (\raisebox{0.3ex}{$\scriptstyle\sim$}14\%), implying many false positives and unclear practical reliability.
\item Best performance relies on expensive flagship LLMs that lead to high monetary cost; cheaper models seem to significantly degrade performance. Some cheaper, open-source models may do better in terms of exploitation success, but internal working effectiveness remains overlooked, casting a shadow on performance.
\item Comparison is limited to Jazzer, lacking evaluation against state-of-the-art analyzers (e.g., IRIS@ICLR'25, RepoAudit@ICML'25, LLMDFA@NeuIPS'24) and stronger fuzzers (e.g., PolyFuzz@USENIX Security'23). While a set of state-of-the-art industry tools is considered, those academic works are overlooked, leaving the technical advancement unclear.
\end{enumerate}

\section{Response to the Meta-Review} 
We appreciate the reviewers' and shepherd's feedback
and offer our perspective on the three concerns.

\PP{Concern 1 (Precision)}%
While our sink filtering precision (7--15.7\%) is admittedly low,
it already matches or exceeds
industry-standard static analysis tools on the same sinks:
CodeQL achieves 6.1\% and SpotBugs 7.6\% (\autoref{t:sa-sota}).
More importantly,
this is a deliberate design choice favoring recall over precision.
Because the downstream agents (exploration and exploitation)
and the fuzzer consume filtered sinks
as their search targets,
missing a true sink means
losing the opportunity
to use agents to accelerate its discovery.
A conservative, recall-oriented filter
therefore minimizes the risk of the agent pipeline
silently discarding a potentially exploitable sink.
Since \sys's end-to-end output consists of
verified proof-of-concept exploits
confirmed by Jazzer's sanitizers,
low precision does not produce false vulnerability reports.
Its practical cost is
increased LLM token consumption on non-exploitable sinks,
reducing overall cost-effectiveness (\autoref{t:agent-filtering}).

\PP{Concern 2 (Cost and Internal Effectiveness)}%
We acknowledge that LLM code reasoning is inherently a black box,
and explaining \emph{why} models succeed at the reasoning level
remains an open challenge for all LLM-based systems.
What our experiments do demonstrate
is that the \emph{framework-level} decomposition is effective:
the ReAct baseline uses the same LLMs
yet performs far below \sys (\autoref{f:tool-coords}),
and the ablation study identifies vulnerabilities
found only through agent-fuzzer synergy (\autoref{f:cpv-matrix}),
confirming that problem decomposition and grounded reasoning---not
raw model capability---drive the gains.
Besides, RQ3 provides per-agent breakdowns
across seven models including open-weight ones
(\autoref{t:agent-filtering},
\autoref{t:agent-exploration},
\autoref{t:agent-exploitation}),
offering both internal observability
and reproducible anchor points
for follow-up research.

\PP{Concern 3 (Comparison Scope)}%
\sys's core contribution is
the first framework that orchestrates
sink knowledge, LLM agents, and coverage-guided fuzzing
to produce dynamic proof-of-concept exploits.
To validate this contribution,
we compare against baselines
spanning each relevant dimension:
static analysis (CodeQL, SpotBugs),
directed fuzzing (Directed Jazzer),
LLM-only reasoning (ReAct agent),
and coverage-guided fuzzing (Jazzer).
The academic tools cited in the concern
(IRIS, RepoAudit, LLMDFA for static detection;
PolyFuzz for multi-language FFI fuzzing)
address complementary problems
and comparing with them
would be a valuable further exploration
that we leave as future work.

\end{document}